\pdfoutput=1

\documentclass[twocolumn]{aastex63}
\usepackage{natbib}
\usepackage{hyperref}
\usepackage{lineno}
\usepackage{amsmath,amssymb}
\usepackage{ulem}

\def\deg{\ifmmode^\circ\else$^\circ$\fi}

\shorttitle{Probing hidden structures in Mon R2 HFS}
\shortauthors{L.~K. Dewangan et al.}

\begin{document}
%



\title{Mon R2: A Hub-Filament System with an Infrared Bubble at the Hub center}

\correspondingauthor{L.~K. Dewangan}
\email{Email: lokeshd@prl.res.in}

\author[0000-0001-6725-0483]{L.~K. Dewangan}
\affiliation{Astronomy \& Astrophysics Division, Physical Research Laboratory, Navrangpura, Ahmedabad 380009, India.}

\author[0000-0001-8812-8460]{N.~K.~Bhadari}
\affiliation{Astronomy \& Astrophysics Division, Physical Research Laboratory, Navrangpura, Ahmedabad 380009, India.}

\author[0000-0002-7367-9355]{A.~K. Maity}
\affiliation{Astronomy \& Astrophysics Division, Physical Research Laboratory, Navrangpura, Ahmedabad 380009, India.}
\affiliation{Indian Institute of Technology Gandhinagar Palaj, Gandhinagar 382355, India.}

\author[0000-0002-7367-9355]{O.~R. Jadhav}
\affiliation{Astronomy \& Astrophysics Division, Physical Research Laboratory, Navrangpura, Ahmedabad 380009, India.}
\affiliation{Indian Institute of Technology Gandhinagar Palaj, Gandhinagar 382355, India.}

\author[0000-0001-5731-3057]{Saurabh Sharma}
\affiliation{Aryabhatta Research Institute of Observational Sciences, Manora Peak, Nainital 263002, India.}

\author[0000-0003-4941-5154]{A. {Haj Ismail}}
\affiliation{College of Humanities and Sciences, Ajman University, 346 Ajman, United Arab Emirates}

\begin{abstract}
A multi-wavelength, multi-scale study of the Mon R2 hub-filament system (HFS) reveals a spiral structure, with the central hub containing more mass than its filaments. ALMA C$^{18}$O(1--0) emission reveals several accreting filaments connected to a molecular ring (size $\sim$0.18 pc $\times$ 0.26 pc).
The molecular ring surrounds the infrared (IR) ring (size $\sim$0.12 pc $\times$ 0.16 pc), which is not usually observed.
The IR ring encircles IR dark regions and a population of embedded near-IR sources, including the massive stars IRS 1 and IRS 2. ALMA HNC(3--2) line data reveal a mirrored B-shaped feature (extent $\sim$19000 AU $\times$ 39000 AU) toward the eastern part of the molecular ring, suggesting expansion at $\sim$2.25 km s$^{-1}$. 
Distinct HNC sub-structures in both redshifted and blueshifted velocity 
components are investigated toward the B-shaped feature. The presence of these braid-like substructures in each velocity component strongly suggests instability in photon-dominated regions. A dusty shell-like feature (extent $\sim$0.04 pc $\times$ 0.07 pc; mass $\sim$7 M$_{\odot}$) hosting IRS~1 is identified in the ALMA 1.14 mm continuum map, centered toward the base of the B-shaped feature. The IR and dense molecular rings are likely shaped by feedback from massive stars, driven by high pressure values between 10$^{-8}$--10$^{-10}$ dynes cm$^{-2}$, observed within a 1 pc range of the B0 ZAMS star powering the ultracompact H\,{\sc ii} region. Overall, these outcomes support that the Mon R2 HFS transitioned from IR-quiet to IR-bright, driven by the interaction between gas accretion and feedback from massive stars.
%
%
%
\end{abstract}
%
%
\section{Introduction}
\label{sec:intro}
The formation process of massive stars  ($>$ 8 M$_{\odot}$) is not yet fully understood \citep[e.g.,][]{Motte+2018,rosen20}.
It is thought that the formation of such stars is intricately linked to accretion through 
filaments in hub-filament systems \citep[HFSs;][]{myers09}. The physical processes driving star formation in HFSs are addressed in the global non-isotropic collapse (GNIC) scenario \citep{Tige+2017,Motte+2018}, the Filaments to Clusters (F2C) scheme \citep{kumar20}, and the inertial inflow model \citep{padoan20}.  
Note that the GNIC scenario incorporates the characteristics of the competitive accretion (CA) model \citep{bonnell01,bonnell04} and the global hierarchical collapse (GHC) model \citep{semadeni09,semadeni17,semadeni19}. According to these existing scenarios \citep[e.g.,][]{Motte+2018,kumar20,padoan20}, in HFSs, gravity-driven and/or turbulence-driven gas motion contributes to mass accumulation in the central hub \citep[see also][]{bhadari23,liu23,yang23}, where the initial seeds of star formation develope and eventually 
grow into massive stars \citep[e.g.,][]{Motte+2018}. However, once massive stars form, they interact with their immediate surroundings through radiative and mechanical feedback. Hence, the interaction between gas accretion through filaments and the feedback from massive stars may significantly shape the morphology of HFSs (including the size of hubs) and influence star formation activities.
In such systems, observationally investigating clues about the origin of massive stars can be extremely challenging, mainly because of the significant impact those massive stars have in the hub. In this relation, a careful study of continuum and molecular line observations across various wavelengths and spatial scales of HFSs will be helpful. The target of this study is the cluster-forming HFS in Monoceros R2/Mon R2 \citep{rayner17,morales19,kumar22}, which is one of the nearest HFSs located at a distance of $\sim$830~pc \citep{racine68,herbst76}. 

In the center of Mon R2, at least five infrared (IR) sources (i.e., IRS~1, IRS~2, IRS~3, IRS~4, and IRS~5) and an embedded cluster have been detected \citep{beckwith76,carpenter97,carpenter08}. The luminosities of IRS~1, IRS~2, IRS~3, IRS~4, and IRS~5 were reported to be $\sim$2000~L$_\odot$, $\sim$100~L$_\odot$, 
$\sim$3000~L$_\odot$, $\sim$100~L$_\odot$, and $\sim$300 L$_\odot$, respectively \citep[see Table~3 in][]{hackwell82}. Using the deep near-IR (NIR) imaging observations, \citet{carpenter97} carried out an extensive study of the dense cluster containing about 500 stars and protostars in the Mon R2 molecular cloud. Using the Hubble Space Telescope (\emph{HST}) NICMOS2 photometric data, further research targeted the fainter, lower-mass populations in the cluster \citep{andersen06}. 

The gas in the Mon R2 molecular cloud is mainly concentrated around the central ultracompact (UC) H\,{\sc ii} region \citep[see a review article by][for more details]{carpenter08}. 
The UCH\,{\sc ii} region has a cometary shape \citep{wood89}, and its maximum continuum brightness reaches 
toward IRS~1 \citep[see Figure~2 in][]{kwon16} that was proposed as the exciting source of the UCH\,{\sc ii} region with a spectral type of B0 ZAMS \citep{downes75,beckwith76,massi85,henning92,fuente10}. 
The UCH\,{\sc ii} region around IRS~1 is surrounded by layers of atomic
and molecular gas (photon-dominated regions (PDRs)) \citep[e.g.,][]{berne09,ginard12}. It is reported that these PDRs exhibit an approximately circular spatial distribution, with a projected thickness ranging from 4$''$ to 6$''$  \citep[e.g.,][]{berne09,pilleri14}. Furthermore, in the direction of the UCH\,{\sc ii} region \citep[age $\sim$10$^{5}$ yr;][]{didelon15}, IR images have unveiled an elliptical ring (or IR ring) enclosing IRS~1 and IRS~2 \citep{beckwith76,andersen06,dewit09}. The position of IRS 1 in the NIR map at 1.65 $\mu$m is about 7$''$ away from the peak position observed in radio continuum map at 6 cm, consistent with the blister H\,{\sc ii} region model described by \citet{massi85}. 
Based on the analysis of spectral line observations (beam size $\sim$23\rlap.{$''$}5--29$''$) obtained with the IRAM-30 m telescope, \citet{morales19} found spiral and ring structures around the source IRS~1 in integrated C$^{18}$O and $^{13}$CO intensity maps, respectively. 
Figure~\ref{fg1}a presents the K-band image at 2.2 $\mu$m overlaid with the radio continuum emission contours at 4.8 GHz, revealing the IR ring, the locations of five IR sources, and the distribution of radio continuum emission. The direction in the figure is presented in Galactic coordinates.

Using high resolution observations, \citet{jimenez13} reported the detection of a new radio recombination line maser object toward IRS~2, while \citet{jimenez20} proposed the distribution of the ionized gas around IRS~2 in a Keplerian circumstellar disk and an expanding wind. IRS~3 and IRS~5 are identified as the youngest and most massive sources, and are not associated with any H\,{\sc ii} region \citep[e.g.,][]{carpenter08}. Both these sources are in the hot core stage \citep{boonman03,dierickx15}. IRS~3 drives a powerful massive outflow \citep{dierickx15,fuente21}, and a collimated CO outflow was also detected toward IRS~5 \citep{dierickx15}. IRS~1, which is powering source of the UCH\,{\sc ii} region, has been considered a more evolved massive star compared to IRS~3 \citep[e.g.,][]{massi85,carpenter08}. 


Several filaments are observed to merge toward the Mon~R2 hub \citep[radius $\sim$0.8 pc;][]{kumar22}, with an accretion rate ranging from 10$^{-3}$ to 10$^{-4}$ M$_{\odot}$ yr$^{-1}$ \citep{morales19}. Based on the detection of the HFS, the GNIC scenario \citep[e.g.,][]{morales19} and the F2C scheme \citep[e.g.,][]{kumar22} have been proposed to explain the observed morphology and star formation activities in Mon~R2. However, the inner environment of the hub, which directly links to the mass inflow process and the origin of IRS sources, is not well understood. In this context, high-resolution molecular line observations (including dense gas and PDR tracers, at scales below 5000 AU) are yet to be examined for the different structures around the IR sources IRS 1 and IRS 2 in Mon~R2.


In order to probe the mass accumulation process and the impact of massive stars formed in the 
Mon~R2 HFS, we study the gas kinematics across different physical scales using the multi-scale and multi-wavelength continuum and line data sets from Atacama Large Millimeter/sub-millimeter Array (ALMA; resolution $\sim$0\rlap.{$''$}77--3\rlap.{$''$}55 or $\sim$639.0~AU--2946.5~AU at a distance of 830 pc). 
This study specifically examines the detailed morphological and kinematical structure of the molecular gas (at different physical scales) associated with the central hub in Mon R2. Additionally, in this context, we also carefully revisit the existing \emph{HST} NIR images. 

Section~\ref{sec:obser} describes the observational data sets utilized in this paper. 
Our observational results are detailed in Section~\ref{sec:data}. 
Section~\ref{sec:disc} discusses the implications of our results toward Mon R2. Finally, Section~\ref{sec:conc} summarizes the conclusions of this work.
\section{Data sets}
\label{sec:obser}
\begin{figure*}
\center
\includegraphics[width=15cm]{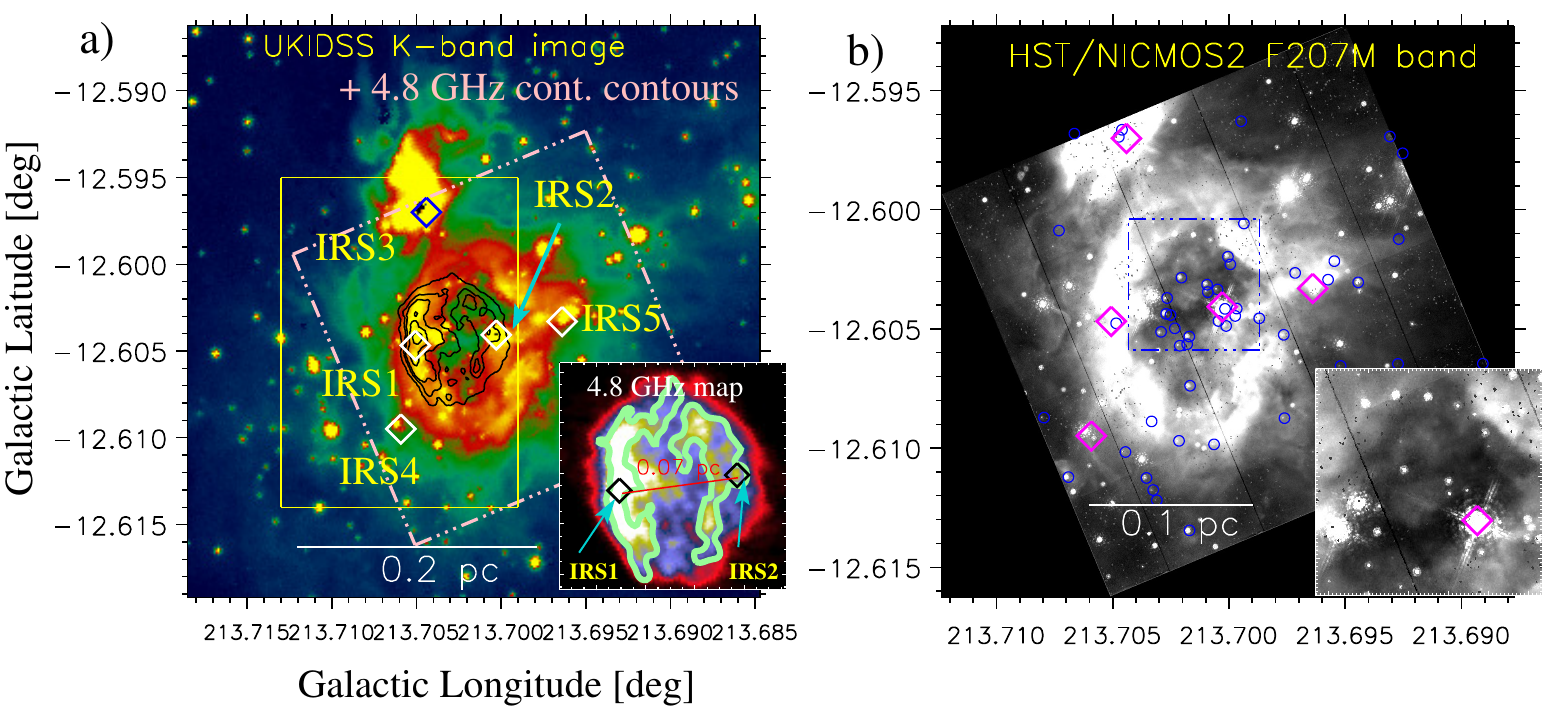}
\includegraphics[width=15cm]{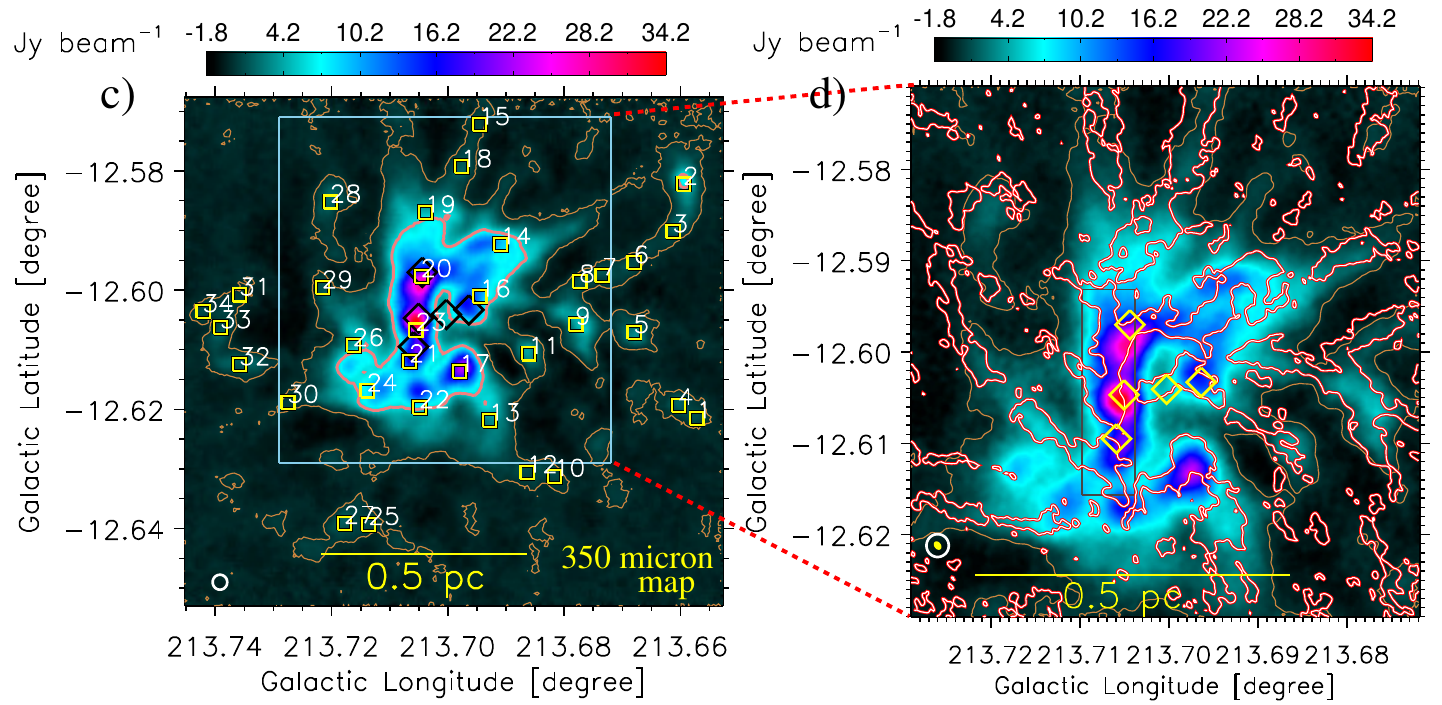}
\caption{a) The UKIDSS K-band image overlaid with the NVAS radio continuum 
emission contours at 4.8 GHz. The radio contours are shown with the levels of 5, 15, 20, 24, 50, and 85 mJy beam$^{-1}$.   
The inset on the bottom right exhibits the area containing the NVAS 4.8 GHz continuum emission. The thick curves (in pale green) show the arc-like features. 
The solid box highlights the area covered by the ALMA band-6 observations, while the dot-dashed box shows the area covered by the \emph{HST}/NICMOS2. 
b) The panel displays the \emph{HST}/NICMOS2 F207M band image overlaid with embedded point-like sources with H $-$ K $\ge$ 2.3 mag (see open circles). 
The inset on the bottom right shows zoomed-in view at the same band.
c) The panel shows the SHARC-II 350 $\mu$m emission map, which is overlaid with the 350 $\mu$m continuum sources (see open squares and also Table~\ref{tab1}). The peru contour at 0.34 Jy beam $^{-1}$ 
and the light coral contour at 6 Jy beam $^{-1}$ are the SHARC-II 350 $\mu$m emission (see text for details).
d) The panel presents a zoomed-in view using the SHARC-II 350 $\mu$m emission map (see box in Figure~\ref{fg1}c) overlaid with the column density ($N$(H$_2$)) contour (in red) at 1.2 $\times$ 10$^{20}$ cm$^{-2}$. 
The area highlighted by the solid box is presented in Figures~\ref{fg10}a--\ref{fg10}e.  In each panel, the scale bar is computed at a distance of 830 pc, and the positions of five IR sources (i.e., IRS~1, IRS~2, IRS~3, IRS~4, and IRS~5) are highlighted by diamonds. The SHARC-II continuum map was
smoothed using a boxcar averaging method with a width of 2 pixels.} 
\label{fg1}
\end{figure*}

In order to probe the dust and gas emission toward the Mon R2 HFS, the science-ready ALMA\footnote[1]{https://almascience.nao.ac.jp/aq/} datasets in bands-3 and 6 were employed. The observations in band-3 (see the solid box in Figure~\ref{fg1}c) cover a large area compared to those 
in band-6 (see the solid box in Figure~\ref{fg1}a). We utilized the primary beam response corrected ALMA datasets given by the ALMA pipeline. The continuum map at 1.14 mm (beam size $\sim$0\rlap.{$''$}91 $\times$ 0\rlap.{$''$}62; 1$\sigma$ $\sim$0.3 mJy~beam$^{-1}$) and molecular lines taken in band~6 with the ALMA (project ID: \#2015.1.00453.S; PI: Asuncion Fuente) were examined. These molecular lines are HNC(3--2) ($\nu_{rest}\sim$271.9811 GHz; beam size $\sim$0\rlap.{$''$}87 $\times$ 0\rlap.{$''$}62), H$^{13}$CN(3--2) ($\nu_{rest}\sim$259.0118 GHz; beam size $\sim$0\rlap.{$''$}91 $\times$ 0\rlap.{$''$}65), CCH(3--2) ($\nu_{rest}\sim$262.00648 GHz; beam size $\sim$0\rlap.{$''$}9 $\times$ 0\rlap.{$''$}63), and CCH(4--3) ($\nu_{rest}\sim$262.00426 GHz; beam size $\sim$0\rlap.{$''$}9 $\times$ 0\rlap.{$''$}63).
The brightness sensitivity of the HNC(3--2), H$^{13}$CN(3--2), CCH(3--2), and CCH(4--3) lines is determined to be $\sim$15, $\sim$15, $\sim$15, and $\sim$12 mJy~beam$^{-1}$, respectively.

Furthermore, we also used the lines C$^{18}$O(1--0) ($\nu_{rest}\sim$109.7822 GHz; beam size $\sim$3\rlap.{$''$}68 $\times$ 2\rlap.{$''$}64) and CS(2--1) ($\nu_{rest}\sim$97.98095 GHz; beam size $\sim$4\rlap.{$''$}14 $\times$ 2\rlap.{$''$}94), which were observed in the ALMA band~3 (project ID: \#2016.1.01144.S; PI name: Trevi{\~n}o-Morales, Sandra). The brightness sensitivity of the C$^{18}$O(1--0) and CS(2--1) lines is computed to be $\sim$30 mJy~beam$^{-1}$and $\sim$20 mJy~beam$^{-1}$, respectively.
Observations for project \#2015.1.00453.S were conducted with the 12m array and C36-2/3 configuration in Cycle 3, and for project \#2016.1.01144.S with the 12m array and C40-1 configuration in Cycle 4. 

The NRAO VLA Archive Survey (NVAS\footnote[2]{http://www.vla.nrao.edu/astro/nvas/}) radio continuum map at 4.8~GHz \citep[beam size $\sim$1\rlap.{$''$}6 $\times$ 1\rlap.{$''$}2;][]{crossley07} was collected. This study also uses the dust continuum map at 350 $\mu$m (resolution $\sim$8\rlap.{$''$}5) and the source catalog at 350 $\mu$m \citep[see][for more details]{merello15}. 
This continuum map was obtained using the Second-generation Submillimeter High Angular Resolution Camera (SHARC-II\footnote[3]{https://irsa.ipac.caltech.edu/data/BOLOCAM\_GPS/overview.html}) facility. 
We also obtained the {\it HST}\footnote[4]{https://archive.stsci.edu/missions-and-data/hst/} F207M image (Proposal ID: 7417; PI: Meyer, Michael R.), which was observed with the \emph{HST}/NICMOS/NIC2 instrument of NIC2 \citep[see also][]{andersen06}. The \emph{HST}/NICMOS/NIC2 photometry of point-like sources in the F160M, F160W, and F207M bands was also collected from \citet{andersen06}. Following the equations given in \citet{andersen06}, the NICMOS/NIC2 photometry was converted to the CIT system. 
The K-band image at 2.2 $\mu$m (resolution $\sim$0\rlap.{$''$}8) was downloaded from the UKIDSS Galactic Plane Survey\footnote[5]{http://wsa.roe.ac.uk/} \citep[GPS;][]{lawrence07}.
\begin{figure*}
\center
\includegraphics[width=13cm]{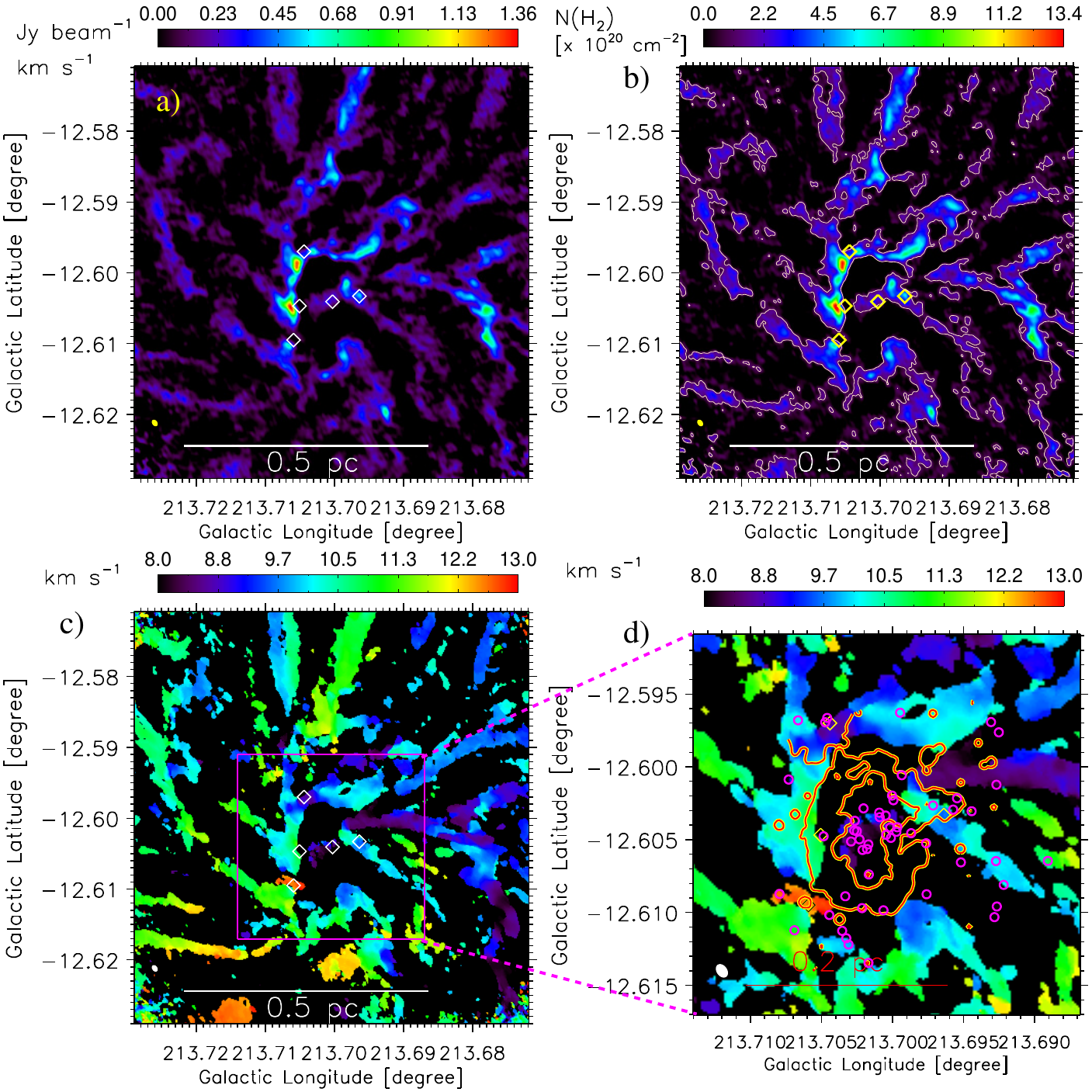}
\caption{a) The panel shows the ALMA C$^{18}$O(1--0) moment-0 map at [8, 13] km s$^{-1}$. 
b) The panel presents the $N(\rm H_2)$ map derived using the ALMA C$^{18}$O(1--0) emission, with the $N(\rm H_2)$ contour level of 1.2 $\times$ 10$^{20}$ cm$^{-2}$. c) The panel displays the ALMA C$^{18}$O(1--0) moment-1 map.  
d) The panel shows the zoomed-in ALMA C$^{18}$O(1--0) moment-1 map 
(see the solid box in Figure~\ref{fg2}c).  
The map is also overlaid with contours (in orange) of the {\it HST}/NICMOS2 F207M band image and the positions of embedded point-like sources with H$-$K $\ge$ 2.3 mag (see open circles and also Figure~\ref{fg1}b). In each panel, the scale bar is computed at a distance of 830 pc, and diamonds are the same as in Figure~\ref{fg1}a.} 
\label{fg2}
\end{figure*}
\begin{figure*}
\center
\includegraphics[width=14cm]{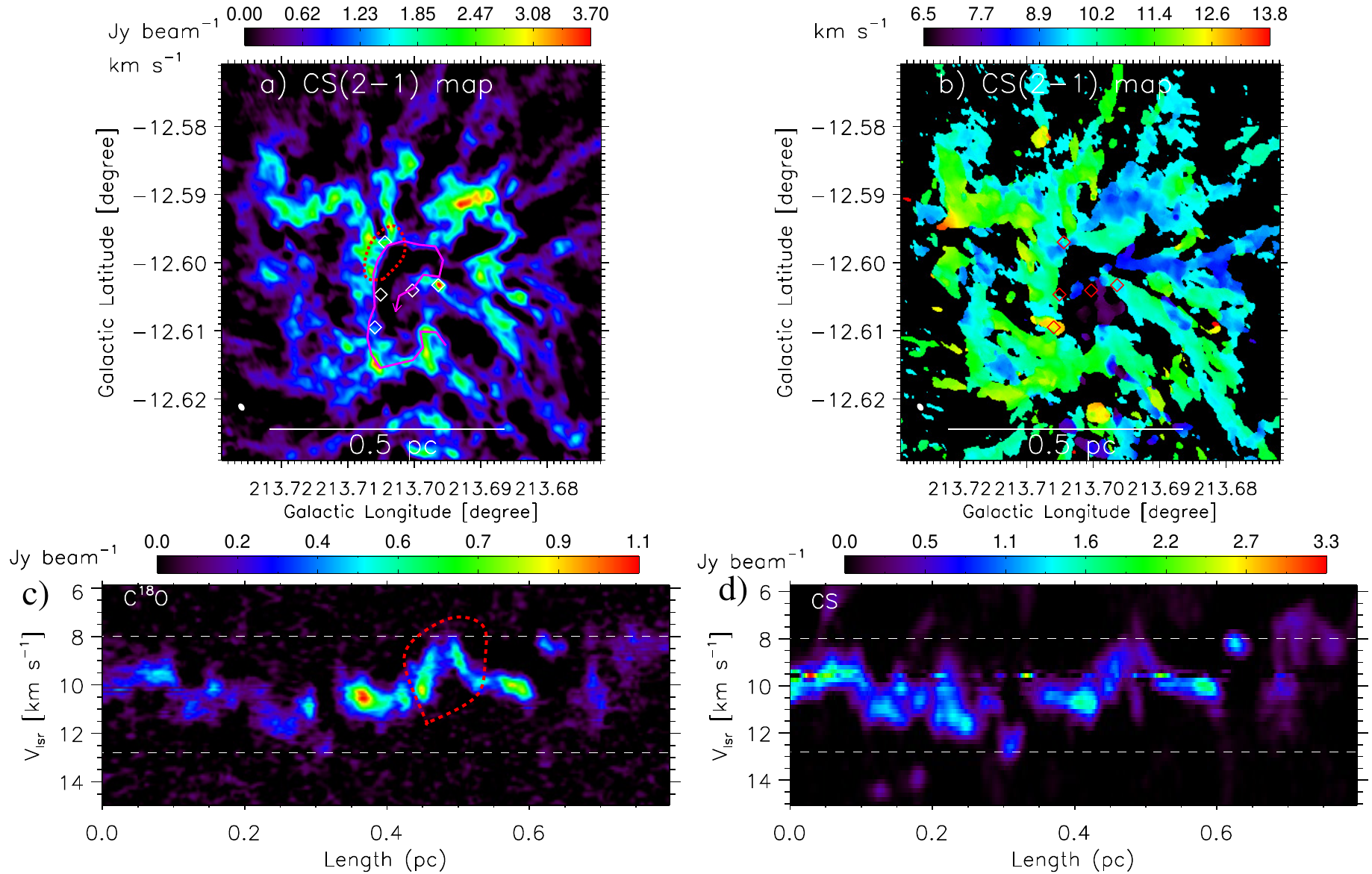}
\caption{a) The panel shows the ALMA CS(2-1) moment-0 map at [5, 14] km s$^{-1}$. 
The molecular ring morphology, visible in the ALMA C$^{18}$O(1--0) and CS(2-1) moment-0 maps, is outlined by a magenta curve based on visual inspection. 
The dashed curve highlights the connection between molecular filaments and molecular ring. b) The panel presents the ALMA CS(2-1) moment-1 map. PV diagram of the c) ALMA C$^{18}$O(1--0); d) ALMA CS(2-1) emission along the curve indicated in Figure~\ref{fg3}a. In panel ``c'', an almost inverted V feature is indicated by the dashed curve. In panels ``a'' and ``b'', the scale bar is computed at a distance of 830 pc, and diamonds are the same as in Figure~\ref{fg1}a.} 
\label{fg3}
\end{figure*}
\section{Results}
\label{sec:data}
\subsection{Physical environment around the central area of Mon R2}
\label{s1sec:d}
Based on several previously published works, it is evident that the central area of Mon R2 is a well-known region of active star formation, home to an embedded cluster and several massive stars. In this section, we study the spatial variations and interactions among the dust, molecular, and ionized features around the central area of Mon R2.
\subsubsection{High resolution NIR maps}
\label{s2sec:d}
In Figures~\ref{fg1}a and~\ref{fg1}b, we display the UKIDSS K-band image and the \emph{HST}/NICMOS2 F207M band image in the direction of an area containing the IRS sources, respectively. 
The IR ring (size $\sim$0.12 pc $\times$ 0.16 pc), which is the most prominent emission structure reported in the literature, is evident in both NIR images.
The {\it Spitzer} 3.6--8 $\mu$m images also display the ring feature \citep[not shown here; see][]{ginard12}, which coincides exactly with the K-band feature. 
%
In Figure~\ref{fg1}a, in the direction of the IR ring, the distribution of the ionized emission traced in the NVAS 4.8 GHz radio continuum map appears to be nearly spherical (extent $\sim$0.12 pc; see also the inset in Figure~\ref{fg1}a). 
However, in the spherical morphology, the majority of the radio continuum emission is primarily concentrated toward the edges containing IRS~1 and IRS~2 rather than at the center. 
The two sources (i.e., IRS~1 and IRS~2) are found to be separated by $\sim$16\rlap.{$''$}6 or $\sim$13780 AU or $\sim$0.07 pc (see the inset in Figure~\ref{fg1}a). 

The center of the IR ring does not show any NIR extended emission features (see Figures~\ref{fg1}a and~\ref{fg1}b). However, an absorption feature is prominent toward the center of the IR ring in the \emph{HST}/NICMOS2 F207M band image, which is also shown in the inset (see Figure~\ref{fg1}b).
Following the work of \citet{andersen06}, we used the photometry of point-like sources in the F160M, F160W, and F207M bands to compute the photometry of these sources in H and K bands (see also Section~\ref{sec:obser}). 
%
Previously, using the positions of point-like sources detected in the NIR maps, \citet{dierickx15} found the concentration of low-mass stars around IRS~1, IRS~2, and IRS~3 in Mon R2 (see Figure~4 in their paper).
To examine the distribution of color-excess or embedded sources, we overlay the point-like sources with H$-$K $\ge$ 2.3 mag on the \emph{HST}/NICMOS2 F207M band image (see open circles in Figure~\ref{fg1}b). This color condition is chosen only to identify sources that are more deeply embedded. 
Some of these embedded sources (excluding IRS~1, IRS~2, and IRS~3) may be low-mass objects \citep[e.g.,][]{andersen06}. 
At least one color-excess source is found toward IRS 1, IRS 2, and IRS 3.
Several embedded sources (including IRS~1 and IRS~2) are detected toward the dark/black regions. 
Considering these outcomes, the significance of the observed dark areas in ongoing star formation processes in Mon R2 cannot be ignored. 
\subsubsection{Sub-millimeter SHARC-II 350 $\mu$m continuum map}
\label{s3sec:d}
%
Previously, the Mon R2 HFS was proposed to resemble a miniature spiral galaxy \citep[e.g.,][]{morales19,kumar22}.
In order to explore the structure of hub, Figure~\ref{fg1}c presents the sub-millimeter SHARC-II 350 $\mu$m continuum map and emission contours (resolution $\sim$8\rlap.{$''$}5) of Mon R2. The SHARC-II map also supports the presence of the spiral-like structure in Mon R2. 
In Figure~\ref{fg1}c, the inner area is indicated by the light coral contour at 6 Jy beam$^{-1}$, where five IRS sources (i.e., IRS 1 -- IRS 5) are located (see diamonds). 
The peak positions of the dust clumps \citep[from][]{merello15} traced in the SHARC-II 350 $\mu$m continuum map are also shown by squares in Figure~\ref{fg1}c (see Table~\ref{tab1} for their positions). \citet{merello15} also provided major axis, minor axis, position angle, deconvolved angular size, and integrated flux density (S$_{\nu}$) of each dust clump. However, the masses of these dust clumps were not reported by \citet{merello15}. Hence, we have estimated the masses of these dust clumps in this paper.

Table~\ref{tab1} lists the mass of each dust clump, which is determined using the following equation (see \citet{hildebrand83} and also equation~3 in \citet{zhang15}):
\begin{equation}
M \, = \, \frac{D^2 \, S_\nu \, R_t}{B_\nu(T_D) \, \kappa_\nu},
\end{equation} 
\noindent where $S_\nu$ represents the total integrated flux (in Jy), 
$D$ is the distance (in kpc), $R_t$ is the gas-to-dust mass ratio, 
$B_\nu$ denotes the Planck function corresponding to a dust temperature $T_D$, 
and $\kappa_\nu$ is the dust absorption coefficient. 
We adopted $\kappa_\nu$ = 1.01\,cm$^2$\,g$^{-1}$ at 350 $\mu$m \citep{ossenkopf94,zhang15}, $R_t$ = 100, $S_\nu$ values from \citet{merello15} (see also Table~\ref{tab1}), and $D$ = 0.83 kpc in this calculation.  \citet{merello15} did not provide temperature values for the dust clumps. In this connection, we use the published {\it Herschel} temperature map of 
Mon R2 by \citet{didelon15}(refer also to Figure~2a in their paper). The average dust temperature in the  inner area of Mon R2, outlined by the light coral contour at 6 Jy beam$^{-1}$ in Figure~\ref{fg1}c, is $\sim$23 K, 
while the areas outside this contour have an average temperature of $\sim$18.5 K. 
The masses of clumps located within the inner area are calculated using $T_D$ $\sim$23~K (see clump IDs with daggers in Table~\ref{tab1}). 
For the other clumps, masses are computed using $T_D$ $\sim$18.5~K. Generally, mass estimation is subject to various uncertainties, including the assumed dust temperature, opacity, and measured flux. As a result, the uncertainty in mass estimates for each continuum source is typically between 10--20\% and can be as high as 50\%. 
%
%
%
The clump masses range from 2.5 M$_{\odot}$ to 717.9 M$_{\odot}$. 
Eleven dust clumps, indicated by daggers in Table~\ref{tab1}, have masses higher than 200 M$_{\odot}$. 

In a similar way, we have also estimated the total mass of the area covered by the light coral contour at 6 Jy beam$^{-1}$ (i.e., inner/central area; see Figure~\ref{fg1}c) to be $\sim$1560 M$_{\odot}$, using $T_D$ $\sim$23~K. Furthermore, the total mass of the area covered by the peru contour at 0.34 Jy beam$^{-1}$  (excluding the area within the light coral contour) is computed to be about 1288 M$_{\odot}$, using $T_D$ $\sim$18.5~K.  
Here, the choice of $T_D$ $\sim$18.5~K over 23~K appears more appropriate \citep[see Figure~2a in][]{didelon15}. Our mass estimates indicate a concentration of more mass toward the center of the spiral structure, where the embedded sources and massive stars are present. However, it is important to note that the mass estimates rely on the strong assumption of constant temperature, which introduces an uncertainty of a few tens of percentage points.  
%

In this analysis, using the IDL's {\it clumpfind} program \citep{williams94}, we estimated the total fluxes for the areas covered by contours at 6 and 0.34 Jy beam$^{-1}$, separately. In the {\it clumpfind} algorithm, contour levels are given as input to define the initial intensity thresholds for clump identification. The algorithm uses these contours to segment regions in the data where the emission exceeds each contour level, iteratively identifying and isolating clumps according to the given thresholds \citep{williams94}. 
%
\subsection{ALMA band-3 molecular line emission}
\label{sec:data2} 
 To study the structure and dynamics of dense molecular gas in Mon R2, we have examined the ALMA C$^{18}$O(1--0) and CS(2--1) lines (beam size $\sim$3\rlap.{$''$}55), which are  commonly used as tracers of dense molecular gas. In this context, the integrated intensity (or moment-0) map at [8, 13] km s$^{-1}$, intensity-weighted velocity (or moment-1) map, and column density ($N$(H$_2$)) map are produced. As indicated earlier, the ALMA band-3 data are available for an area highlighted by the solid box in Figure~\ref{fg1}c. 
The $N$(H$_2$) map is generated using the C$^{18}$O(1--0) integrated intensity map at [8, 13] km s$^{-1}$. Assuming that the emission is optically thin and the gas is in local thermodynamic equilibrium, an expression from \citet{Mangum_Shirley_2015} connects $N$(C$^{18}$O) to the integrated intensity value \citep[see Equation 1 in][]{Dewangan_2019}. The main beam filling factor and cosmic background temperature are adopted to be 1 and 2.73 K, respectively. The column density map is computed under the approximation of a uniform excitation temperature. The dust temperature was employed as a proxy for the C$^{18}$O excitation temperature, using average values of 18.5~K and 23~K (refer to Section~\ref{s3sec:d} for justification).   
The $N$(C$^{18}$O) values are converted to $N$(H$_2$) using the column-density ratio $N$(C$^{18}$O)/$N$(H$_2$) = 1.7 $\times$ 10$^{-7}$, as reported by \citet{Frerking_1982}. 
In this paper, we have only presented the $N$(H$_2$) map computed with an excitation temperature of 23~K. Dividing this $N$(H$_2$) map by a factor of 1.15 produces the equivalent $N$(H$_2$) map computed with an excitation temperature of 18.5 K. 
Generally, the column density can have uncertainties on the order of a few \citep[e.g.,][]{Frerking_1982}. 
These uncertainties arise from factors such as assumptions about optical depth, inaccuracies in excitation temperature, variations in abundance ratios, line-of-sight overlaps, instrumental noise, and simplified radiative transfer assumptions. 
%

Figure~\ref{fg1}d shows an overlay of the $N$(H$_2$) contour at 1.2 $\times$ 10$^{20}$ cm$^{-2}$ on the SHARC-II 350 $\mu$m emission map (see the solid box in Figure~\ref{fg1}c), allowing us to examine the structures traced in both the molecular gas and the dust continuum map. 
The ALMA C$^{18}$O(1--0) map has better spatial resolution compared to the SHARC-II map. In the $N$(H$_2$) map, filamentary structures, that extend in all directions, seem to present beyond the observed area. The $N$(H$_2$) map confirms that the Mon R2 HFS resembles the appearance of a tiny spiral galaxy.  
In Figures~\ref{fg2}a,~\ref{fg2}b, and~\ref{fg2}c, the moment-0 map, $N$(H$_2$) map, and moment-1 map are presented, respectively. The column density map is also overlaid with the $N$(H$_2$) contour at 1.2 $\times$ 10$^{20}$ cm$^{-2}$ (see also Figure~\ref{fg1}d). The intensity and column density maps show the existence of a molecular ring feature (approximately $\sim$0.18 pc $\times$ 0.26 pc in size), with its different parts seemingly interconnected by molecular filaments in every direction. The molecular ring exhibits an elliptical morphology, and its south-west portion is broken. 
Additionally, its central area is not fully compact. Moment-1 map reveals the velocity variations toward the molecular ring and the molecular filaments. 
Figure~\ref{fg2}d displays the zoomed-in view of moment-1 map around the area containing the IR ring, which is indicated by the contours of the {\it HST}/NICMOS2 F207M band image. Interestingly, the molecular ring almost surrounds the IR ring and the spherical structure traced in the NVAS 4.8 GHz radio continuum map (see Figures~\ref{fg2} and~\ref{fg3}). Furthermore, the ALMA C$^{18}$O emission (around 9 km s$^{-1}$) is traced toward the central part of the molecular ring, where the point-like sources with H$-$K $\ge$ 2.3 mag are found (see also Figure~\ref{fg1}b). 
Figures~\ref{fg3}a and~\ref{fg3}b show the moment-0 map at [5, 14] km s$^{-1}$ and moment-1 map of the CS(2--1) emission, respectively. We find that the CS(2--1) emission displays similar features to those observed in the C$^{18}$O emission. 
Both of these molecular emissions are detected toward all IR sources except IRS~1. The molecular gas having different velocities are traced toward the molecular ring and the molecular filaments, which can be noticed from the moment-1 maps presented in 
Figures~\ref{fg2}c and~\ref{fg3}b. 


Figures~\ref{fg3}c and~\ref{fg3}d display position-velocity (PV) diagrams of the C$^{18}$O and CS emissions along the curve, respectively. This curve is chosen in the direction of the molecular ring and its central region (see the curve in Figure~\ref{fg3}a). Both the PV diagrams show the variations in the velocity. 
Additionally, an almost inverted V-shaped velocity feature, displaying a significant velocity gradient (i.e. $\sim$3 km s$^{-1}$ in $\sim$0.1 pc), is evident in both PV diagrams (see the dashed curve in Figure~\ref{fg3}c). 
This velocity feature is observed in the direction of one of the junctions where the molecular filaments converge toward the molecular ring (see the dashed curve in Figure~\ref{fg3}a). 
The observed velocity gradient may highlight the complexity of the gas movement from the filaments to the molecular ring. 
These areas seem to provide crucial insights into the dynamic interactions and connectivity within the molecular configuration. 

In Sections~\ref{subsec:data3a} and~\ref{subsec:data3b}, we have also examined the ALMA band-6 continuum map at 1.14 mm and 
molecular lines (beam size $\sim$0\rlap.{$''$}77), respectively. 
It is important to note that the ALMA band-6 data sets offer better spatial resolution compared to the ALMA band-3 maps. However, the observed area in the band-6 is limited (see the solid box in Figure~\ref{fg1}a). 
\begin{figure*}
\center
\includegraphics[width=\textwidth]{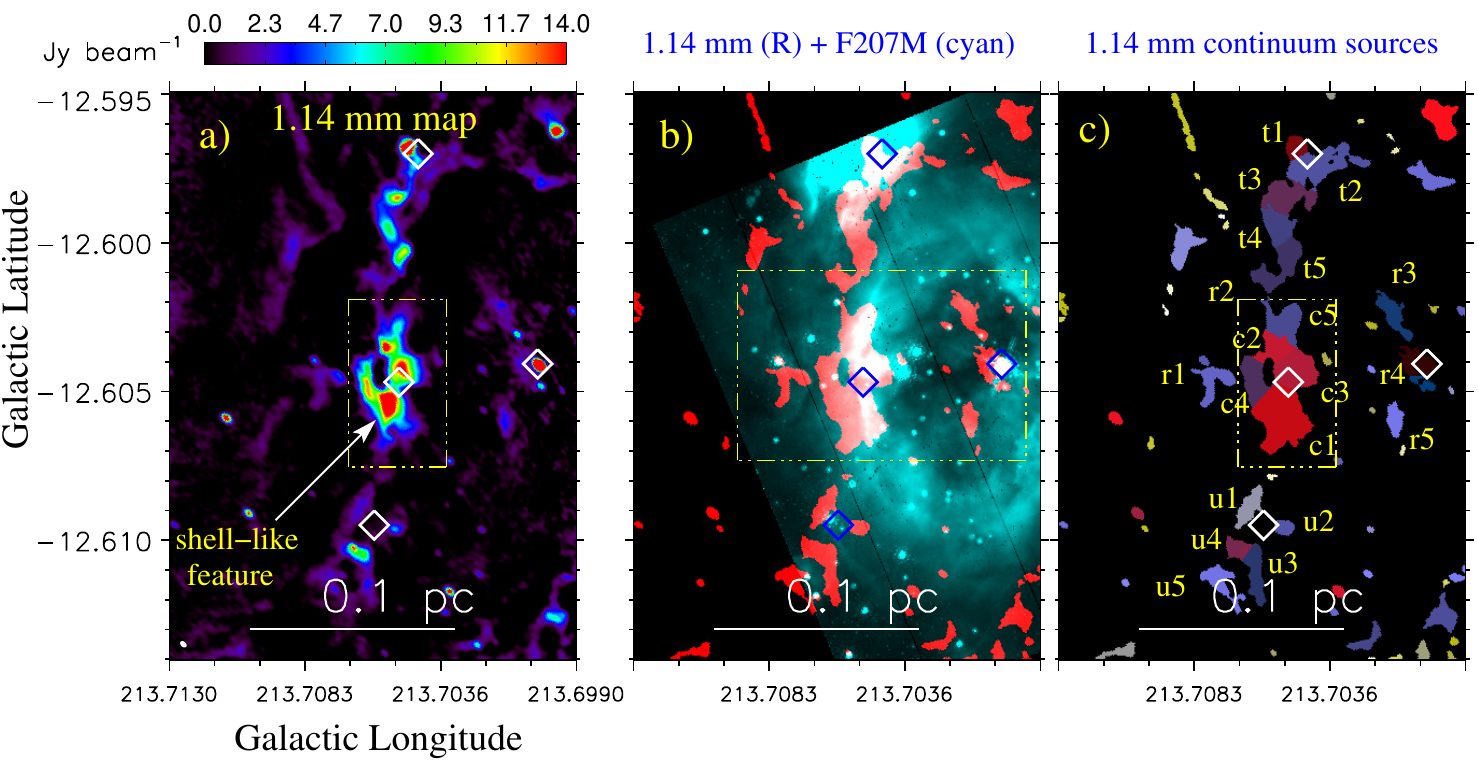}
\caption{a) ALMA 1.14 mm dust continuum map (see the solid box in Figure~\ref{fg1}a). 
b) The panel displays a two-color composite map produced using the ALMA 1.14 mm dust continuum map (in red) and the {\it HST}/NICMOS2 F207M band image (in cyan).  
c) The panel shows a clumpfind decomposition of the ALMA 1.14 mm dust 
continuum emission, highlighting the spatial extension of several continuum sources (i.e., t1--t4, r1--r5, c1--c5, and u1--u5; see Table~\ref{tab2}).  
In each panel, the scale bar is drawn at a distance of 830 pc and diamonds are the same as in Figure~\ref{fg1}a.}
\label{fg4}
\end{figure*}
\begin{figure}
\center
\includegraphics[width=8.5 cm]{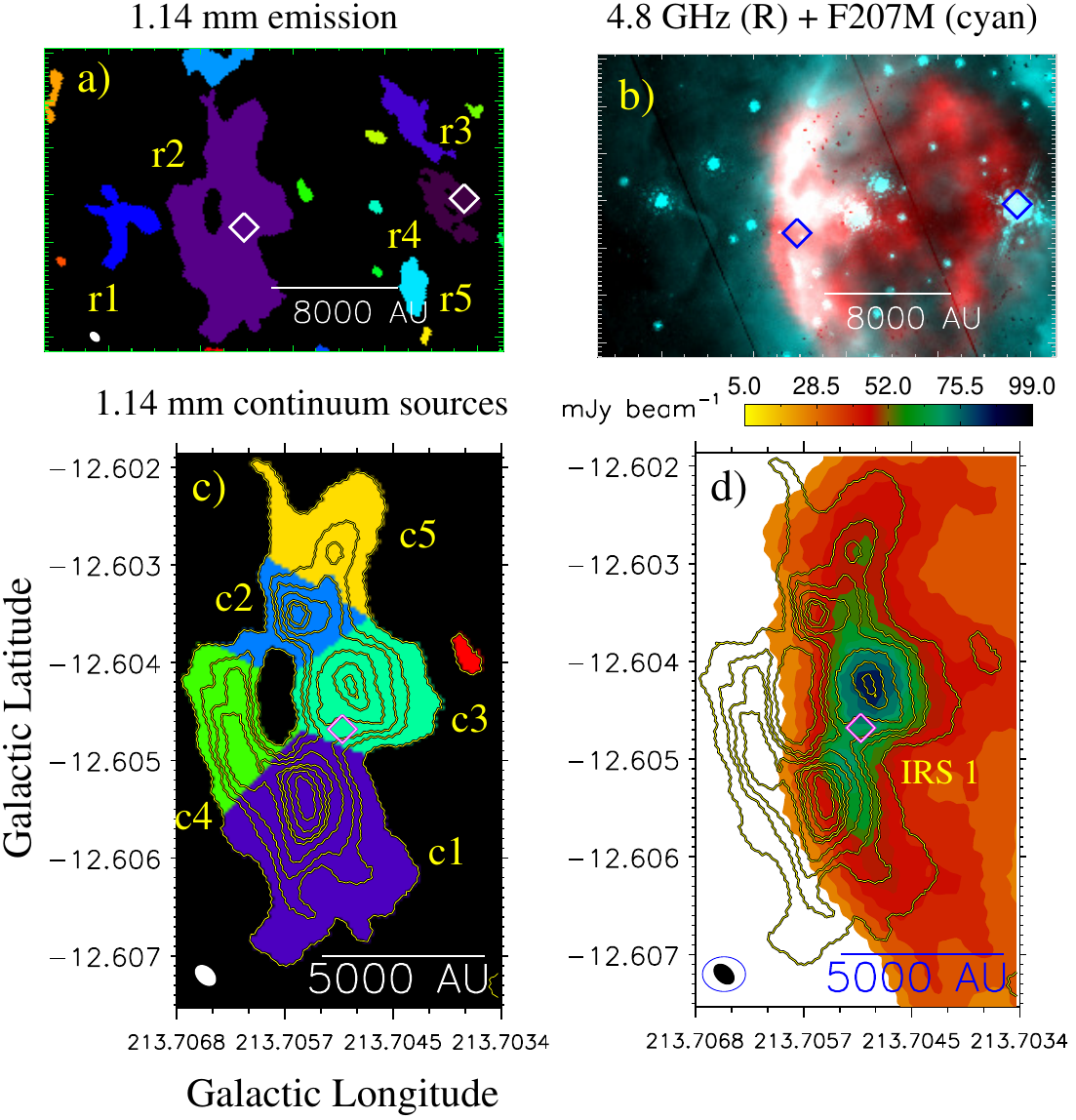}
\caption{a) Clumpfind decomposition of the ALMA 1.14 mm dust 
continuum emission (see the dot-dashed box in Figure~\ref{fg4}b), displaying the spatial extension of at least five 
continuum sources (i.e., r1, r2, r3, r4, and r5; see Figure~\ref{fg4}c and also Table~\ref{tab2}). b) Two-color composite map produced using 
the NVAS 4.8 GHz continuum map (in red) and 
the {\it HST}/NICMOS2 F207M band image (in cyan; see the dot-dashed box in Figure~\ref{fg4}b). c) Clumpfind decomposition of the ALMA 1.14 mm dust 
continuum emission in the direction of continuum source ``r2'' (see the dot-dashed box in Figure~\ref{fg4}c and also Figure~\ref{fg4z}a), 
which is also overlaid with the 1.14 mm continuum emission contours. 
The contours are plotted with the levels of 1.32, 4.24, 6.5, 9.85, 13.25, 15.0, 19.0, and 24.31 
mJy beam$^{-1}$. The spatial extension of five compact continuum 
sources (i.e., c1, c2, c3, c4, and c5; see Figure~\ref{fg4}c and also Table~\ref{tab2}) 
is indicated by the five different colors. d) Overlay of the 1.14 mm continuum emission contours on the NVAS 4.8 GHz continuum map and the contours are the same as in Figure~\ref{fg4z}c. Beam sizes of the 1.14 mm dust continuum map (open ellipse) and the NVAS map (filled ellipse) are shown in the bottom-left corner. 
In each panel, the scale bar is drawn at a distance of 830 pc and diamonds are the same as in Figure~\ref{fg1}a.}
\label{fg4z}
\end{figure}
\begin{figure}
\center
\includegraphics[width=9cm]{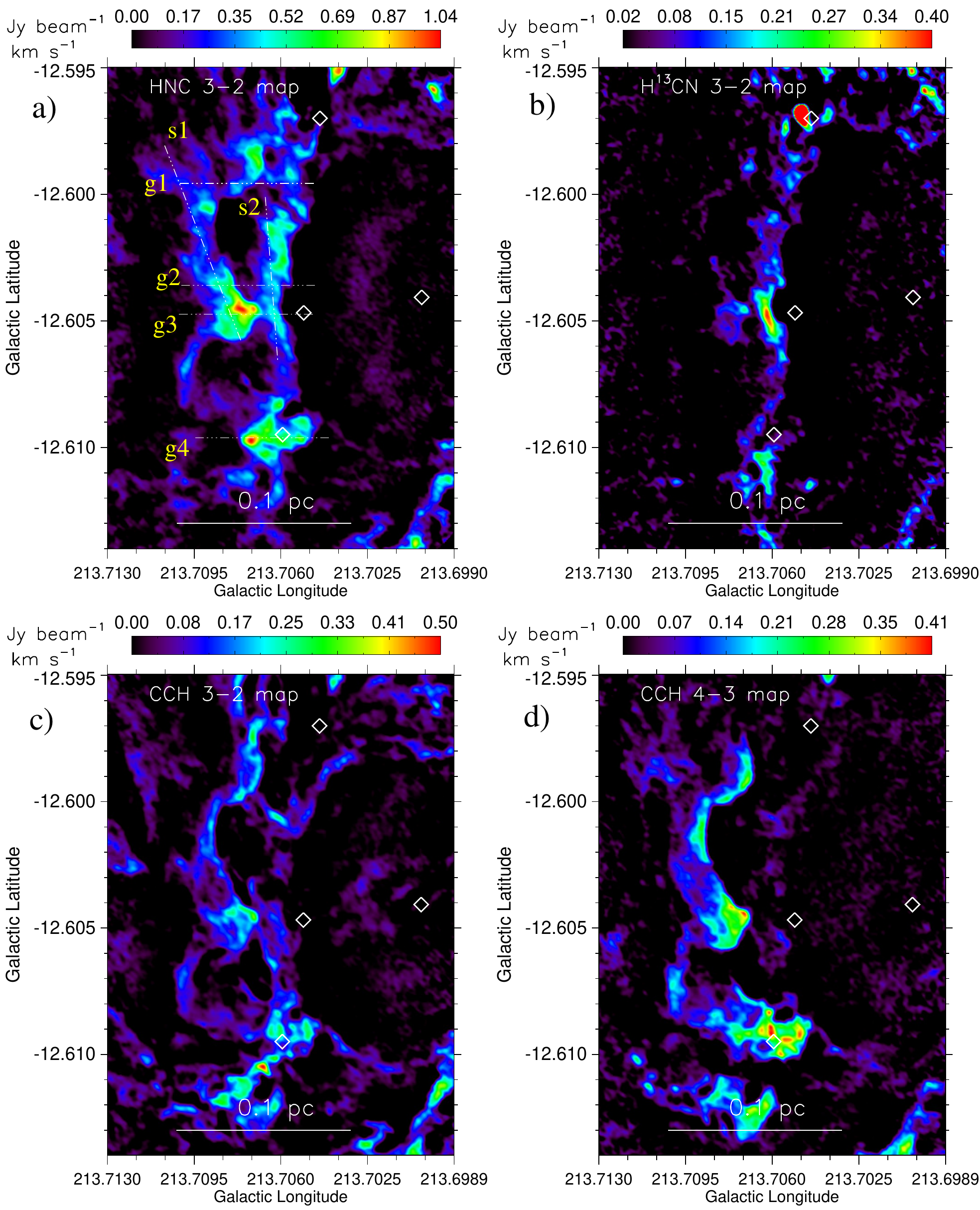}
\includegraphics[width=9cm]{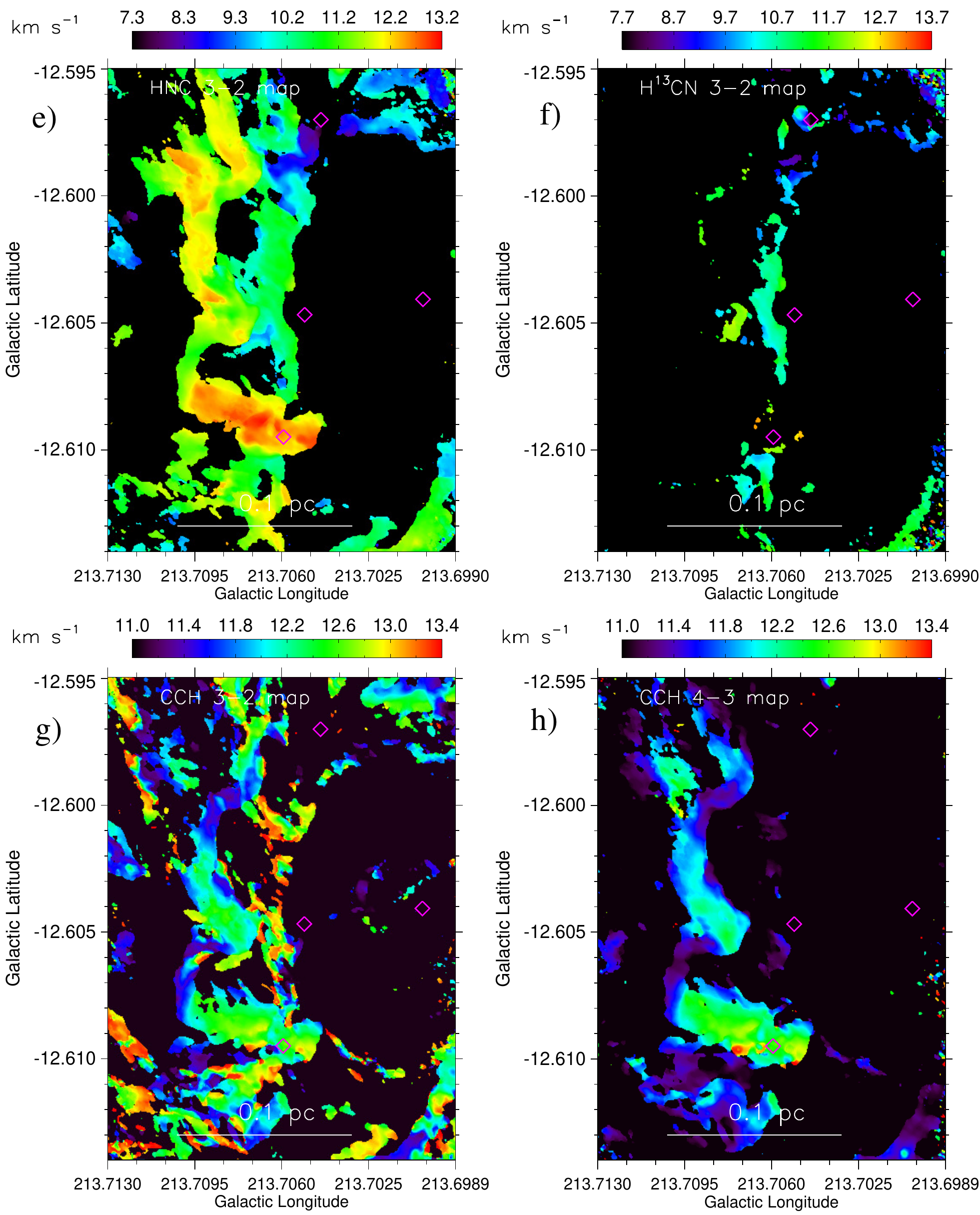}
\caption{
(a--d) Moment-0 maps and (e--h) moment-1 maps of ALMA HNC(3--2), H$^{13}$CN(3--2), CCH(3--2), and CCH(4--3) emission, respectively. In each panel, diamonds are the same as in Figure~\ref{fg1}a.}
\label{fg5}
\end{figure}
%
%
\begin{figure*}
\center
\includegraphics[width=10cm]{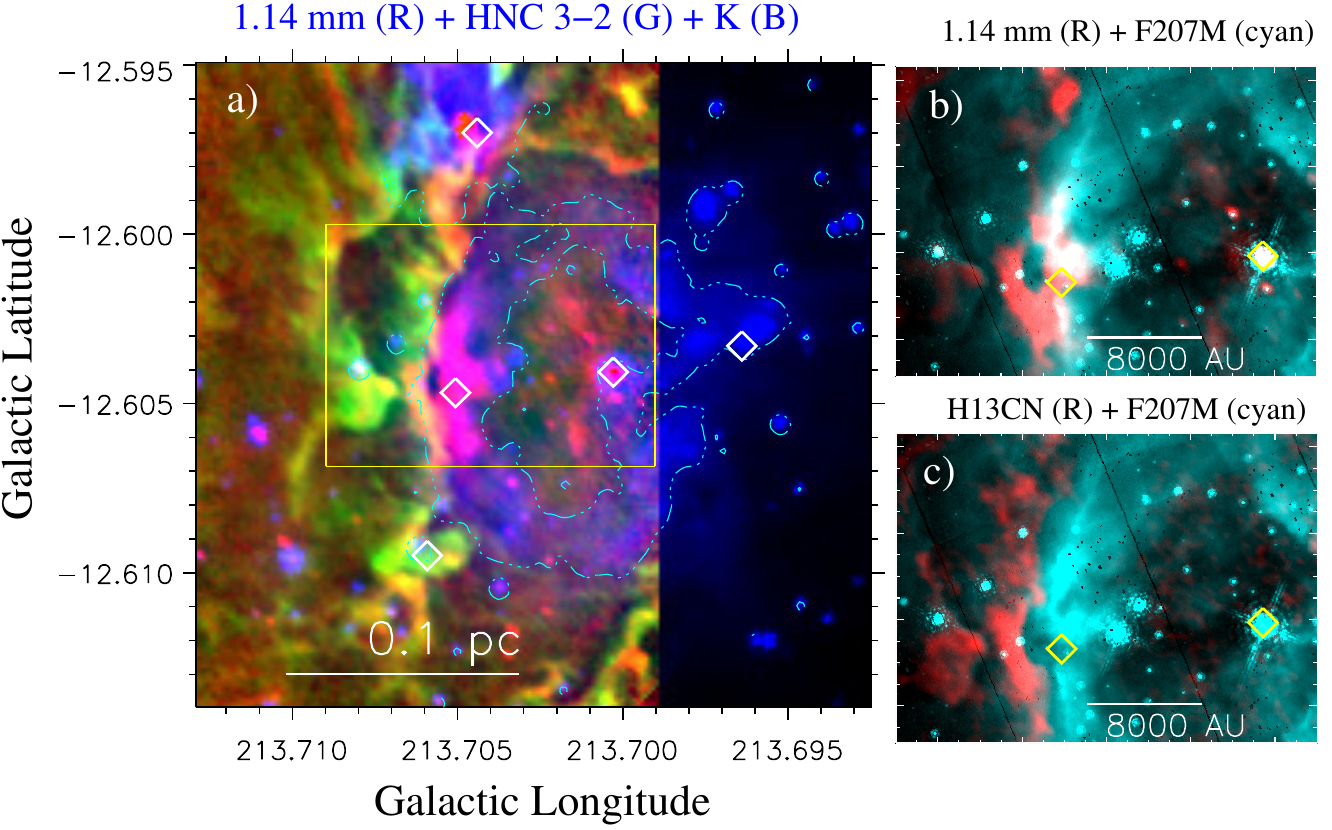}
\includegraphics[width=6 cm]{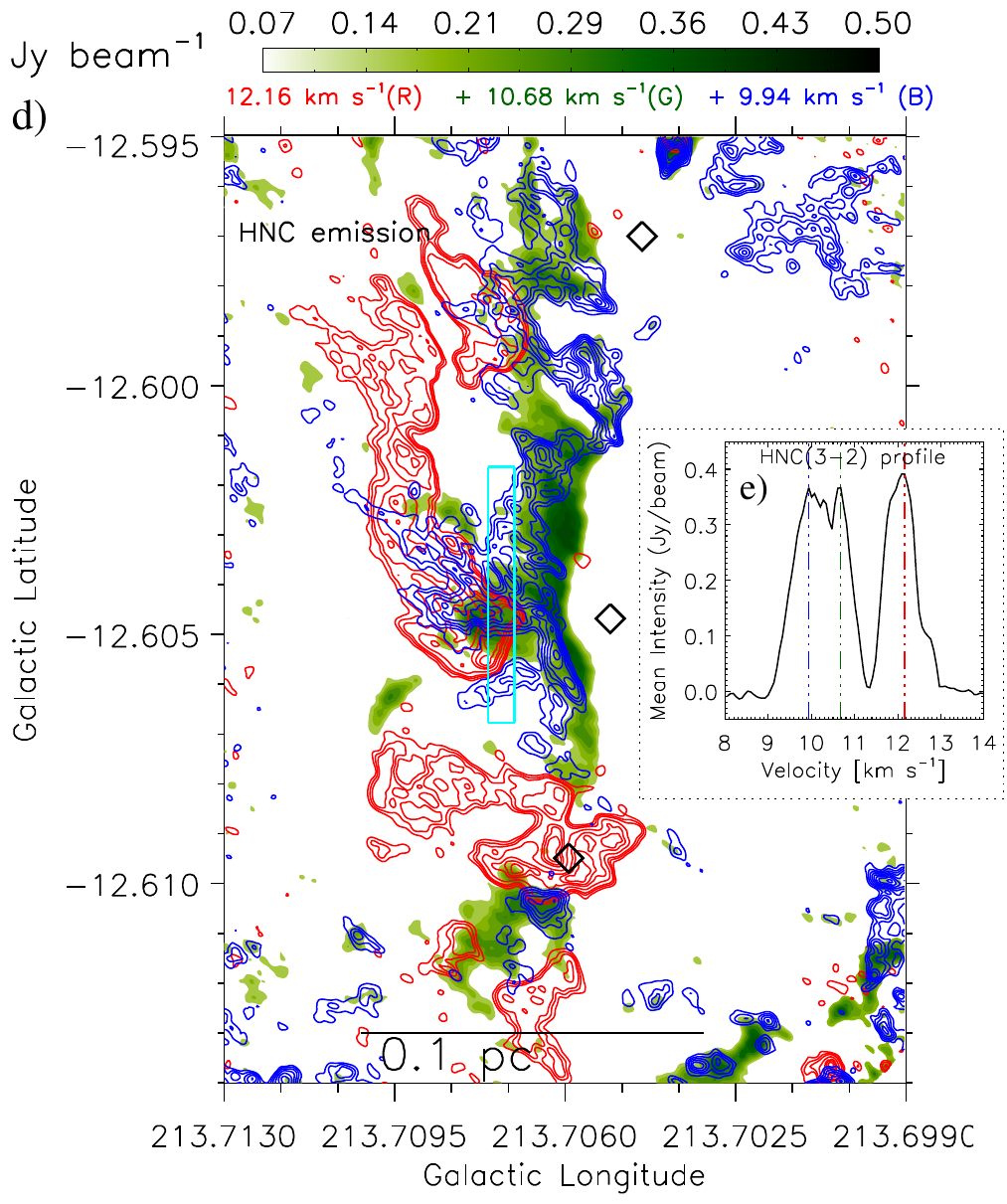}
\caption{a) The panel displays a three-color composite map produced using the ALMA 1.14 mm dust continuum map (in red), the HNC(3--2) moment-0 map (in green), 
and the UKIDSS K-band image (in blue). The dot-dashed curve shows the location of 
IR ring traced in the {\it HST}/NICMOS2 F207M band image. 
b) The panel shows a two-color composite map made using the ALMA 1.14 mm dust continuum map (in red) and the {\it HST}/NICMOS2 F207M band image (in cyan; see a solid box in Figure~\ref{fg7}a). 
c) The panel presents a two-color composite map produced using the H$^{13}$CN(3--2) moment-0 map (in red) and the {\it HST}/NICMOS2 F207M band image (in cyan). 
d) The panel displays the ALMA HNC(3--2) filled contour map at 10.68 km s$^{-1}$ (in green) 
overlaid with contours of HNC(3--2) emission maps at 12.16 km s$^{-1}$ (in red) and 9.94 km s$^{-1}$ (in blue). 
e) Average HNC(3--2) emission profile in the direction of an area highlighted by the rectangle box (in cyan) in Figure~\ref{fg7}d. 
Contours of the HNC(3--2) map at 12.16 km s$^{-1}$ are plotted with the levels of 
0.58 $\times$ [0.12, 0.21, 0.3, 0.5, 0.7, 0.8, 0.9] Jy beam$^{-1}$, 
while the HNC(3--2) emission contours at 9.94 km s$^{-1}$ are shown with the levels of 0.49 $\times$ [0.12, 0.21, 0.3, 0.4, 0.5, 0.6, 0.7, 0.8, 0.9] Jy beam$^{-1}$. 
In each panel, diamonds are the same as in Figure~\ref{fg1}a and the scale bar is produced at a distance of 830 pc.} 
\label{fg7}
\end{figure*}
\subsection{ALMA band-6 continuum emission}
\label{subsec:data3a} 
Figure~\ref{fg4}a displays the ALMA 1.14 mm dust continuum map, revealing a prominent filament (extent $\sim$0.24 pc) containing several dust condensations. Extended dust emissions are detected on both the eastern and western sides of the filament. 
%
%
Interestingly, a small-scale shell-like feature (extent $\sim$0.04 pc $\times$ 0.07 pc; see the dot-dashed box in Figure~\ref{fg4}a) hosting IRS~1 is visible in the ALMA 1.14 mm dust continuum map. Several dust condensations are observed along the edges of 
the shell-like feature, but no dust continuum emission is evident in its central 
area (see Figures~\ref{fg4}a,~\ref{fg4}c, and~\ref{fg4z}c).
%
Previously, the Submillimeter Array (SMA) 1.3 mm continuum emission was observed toward IRS~1, but it was not detected in the 0.85 mm continuum image \citep{dierickx15}. 
However, the shell-like feature containing IRS~1 is not found in the SMA 1.3 mm continuum map \citep[see Figure~1 in][]{dierickx15}. 
In Figure~\ref{fg4}b, we present a two-color composite map generated using the ALMA 1.14 mm dust continuum map (in red) and the {\it HST}/NICMOS2 F207M band image (in cyan), enabling us to compare the distribution of 1.14 mm dust emission with the IR ring. 
 
 In order to identify continuum sources in the ALMA 1.14 mm continuum map, the contours at [1.2, 3.8, 6.3, 8.8, 11.3,13.8, 16.2, 18.7, 21.2, 23.7, 26.2, 28.7] mJy beam$^{-1}$ were given as input to 
 the IDL's {\it clumpfind} program. 
 Figure~\ref{fg4}c displays the spatial extension of the ALMA 1.14 mm continuum sources, which are labeled as ``t1--t5'', ``r1--r5'', ``c1--c5'' and ``u1--u5''.  
Note that the boundary of the continuum source ``r2'' outlines the small-scale shell-like feature hosting IRS~1, traced by the contour at 1.2 mJy beam$^{-1}$. 
The continuum source ``r4'' hosts IRS~2, while IRS~3 seems to be associated with the continuum source ``t1''. 
 %
 
In the direction of the center and one of the edges of the IR ring (see the dot-dashed box in Figure~\ref{fg4}b),  
Figure~\ref{fg4z}a displays the spatial extension of continuum sources, which are designated as r1, r2, r3, r4, and r5 (see also Figure~\ref{fg4}c).  
Figure~\ref{fg4z}b shows a two-color composite map produced using 
the NVAS 4.8 GHz continuum map (in red) and the {\it HST}/NICMOS2 F207M band image (in cyan), and also enables us to examine the ionized emission toward the center and one of the edges of the IR ring. 
Figure~\ref{fg4z}c shows the clumpfind decomposition of the ALMA 1.14 mm dust continuum emission in the direction of source ``r2'' (or shell-like feature), which is also overlaid with the 1.14 mm continuum emission contours. The spatial extension of five compact continuum 
sources (``c1--c5'') within the shell-like feature is presented  (see also Figure~\ref{fg4}c). 
In Figure~\ref{fg4z}d, we show the overlay of the 1.14 mm continuum emission contours on the NVAS 4.8 GHz continuum map toward the shell-like feature, displaying the presence of the ionized emission toward all the compact continuum sources except ``c4'' (i.e., c1--c3 and c5). The IR source IRS~1 and the radio continuum peak emission are associated with the compact continuum source ``c3'', which is located at one of the edges of the IR ring. 
%
%

Using the {\it clumpfind} program, we also computed the total flux, the FWHM not corrected for beam size for the x-axis (i.e., FWHM$_{x}$), and for the y-axis (i.e., FWHM$_{y}$) of each identified source. Table~\ref{tab2} provides the positions, fluxes, deconvolved FWHM$_{x}$ \& FWHM$_{y}$, and masses of all the continuum sources highlighted in Figure~\ref{fg4}c. Following Equation~1, masses are computed using $R_t$ = 100, $T_D$ = 23~K, $\kappa_\nu$ = 1.14\,cm$^2$\,g$^{-1}$ at 1.14 mm \citep{enoch08,bally10,dewangan16}. The mass ranges of continuum sources ``t1--t5'', ``r1--r5'', ``c1--c5'', and ``u1--u5'' are estimated to be 
[0.7, 1], [0.3, 7.0], [0.5, 2.9], and [0.1, 0.4] M$_{\odot}$, respectively. 
The mass of the continuum source, ``r2'' (or shell-like feature), is about 7 M$_{\odot}$, containing the sources``c1--c5''. 
As metioned earlier, the mass estimate for each continuum source often carries an uncertainty of approximately 20\%, which can increase to as much as 
50\% if the temperature is poorly defined. 
\subsection{ALMA band-6 molecular line emission}
\label{subsec:data3b} 
\subsubsection{Mirrored B-like feature}
\label{zsubsec:data3i} 
This section focuses on the molecular lines observed in ALMA band-6, and the lines are HNC(3--2), H$^{13}$CN(3--2), CCH(3--2), and CCH(4--3). The H$^{13}$CN(3--2) line is a recognized indicator of dense molecular gas, while the HNC(3--2), CCH(3--2), and CCH(4--3) lines  serve as tracers of PDRs in H\,{\sc ii} regions. 
Furthermore, the HNC(3--2) line is useful for tracing dense molecular gas close to the UV-illuminated surfaces, whereas CCH, being a radical, is typically observed in the outer regions of PDRs, close to their interface with H\,{\sc ii} regions.   
We have examined the HNC(3--2) emission at [7, 14.2] km s$^{-1}$, the H$^{13}$CN(3--2) emission at [7.7, 13.8] km s$^{-1}$, the CCH(3--2) emission at [11,13.4] km s$^{-1}$, and the CCH(4--3) emission at [11, 13.9] km s$^{-1}$ toward the area outlined by the solid box in Figure~\ref{fg1}a. 

In Figures~\ref{fg5}a,~\ref{fg5}b,~\ref{fg5}c, and~\ref{fg5}d, we present the moment-0 maps of the HNC(3--2), H$^{13}$CN(3--2), CCH(3--2), and CCH(4--3) emission, respectively. A mirrored B-like feature (i.e., \(\reflectbox{B}\)) is apparent as the most prominent molecular structure (extent $\sim$19000 AU $\times$ 39000 AU) in the moment-0 map of the HNC(3--2) emission. Due to higher resolution of the ALMA band-6 data sets, our study also enables us to examine the gas distribution toward the outer region, central part, and base segment of the mirrored B-like structure. 
The mirrored B-like feature is also evident in the CCH(3--2) and CCH(4--3) moment-0 maps. 
 However, the CCH(3--2) and CCH(4--3) emissions are found more intense toward the outer region and the central part of the mirrored B-like structure. The H$^{13}$CN(3--2) moment-0 map mainly traces the base segment and central part of the mirrored B-like feature, along with a molecular condensation of higher intensity toward IRS~3. We find noticeable molecular emissions around IRS~4 in the HNC(3--2), CCH(3--2), and CCH(4--3) emission maps. No molecular emission is observed toward IRS~1. In Figures~\ref{fg5}e,~\ref{fg5}f,~\ref{fg5}g, and~\ref{fg5}h, we display the HNC(3--2), H$^{13}$CN(3--2), CCH(3--2), and CCH(4--3) moment-1 maps, respectively, showing the presence of noticeable velocity variations toward mirrored B-like feature.

Figure~\ref{fg7}a shows a three-color composite map produced using the ALMA 1.14 mm dust continuum map (in red), the HNC(3--2) moment-0 map (in green), and the UKIDSS K-band image (in blue), where the location of the IR ring is highlighted by the dot-dashed curve. It allows us to study the distribution of the HNC(3--2) emission and the 1.14 mm dust continuum emission with respect to the IR ring. We find that the ALMA 1.14 mm dust continuum emission is detected across the mirrored B-like structure, and the continuum source ``r1'' is situated at its central part (see also Figure~\ref{fg4}). 
In Figure~\ref{fg7}b, we show a two-color composite map made using the ALMA 1.14 mm dust continuum map (in red) and the {\it HST}/NICMOS2 F207M band image (in cyan; see a solid box in Figure~\ref{fg7}a). In Figure~\ref{fg7}c, we present a two-color composite map produced using the H$^{13}$CN(3--2) moment-0 map (in red) and the {\it HST}/NICMOS2 F207M band image (in cyan). 
The central part of the IR ring is not depicted in the HNC(3--2) and H$^{13}$CN(3--2) maps (see Figures~\ref{fg7}b and~\ref{fg7}c). The shell-like feature is divided into two distinct regions: one half, containing IRS~1, shows radio continuum emission but lacks molecular emission, while the opposite half is associated with molecular emission and is part of the base segment of the mirrored B-like structure, which does not exhibit any radio continuum emission. 
The absence of molecular gas detection in some regions of the dust shell prevents a complete study of the gas distribution throughout the entire shell-like structure.
Additionally, the ALMA band-6 data sets, despite having limited coverage, also indicate that the molecular emission surrounds the IR ring. 
We have also generated moment-2 map or velocity dispersion map using the HNC(3--2) emission (not shown here), 
which shows higher values of velocity dispersion ($\sim$2.2--3.2 km s$^{-1}$) toward the central part of the mirrored B-like structure and IRS~4. 
\begin{figure*}
\center
\includegraphics[width=12cm]{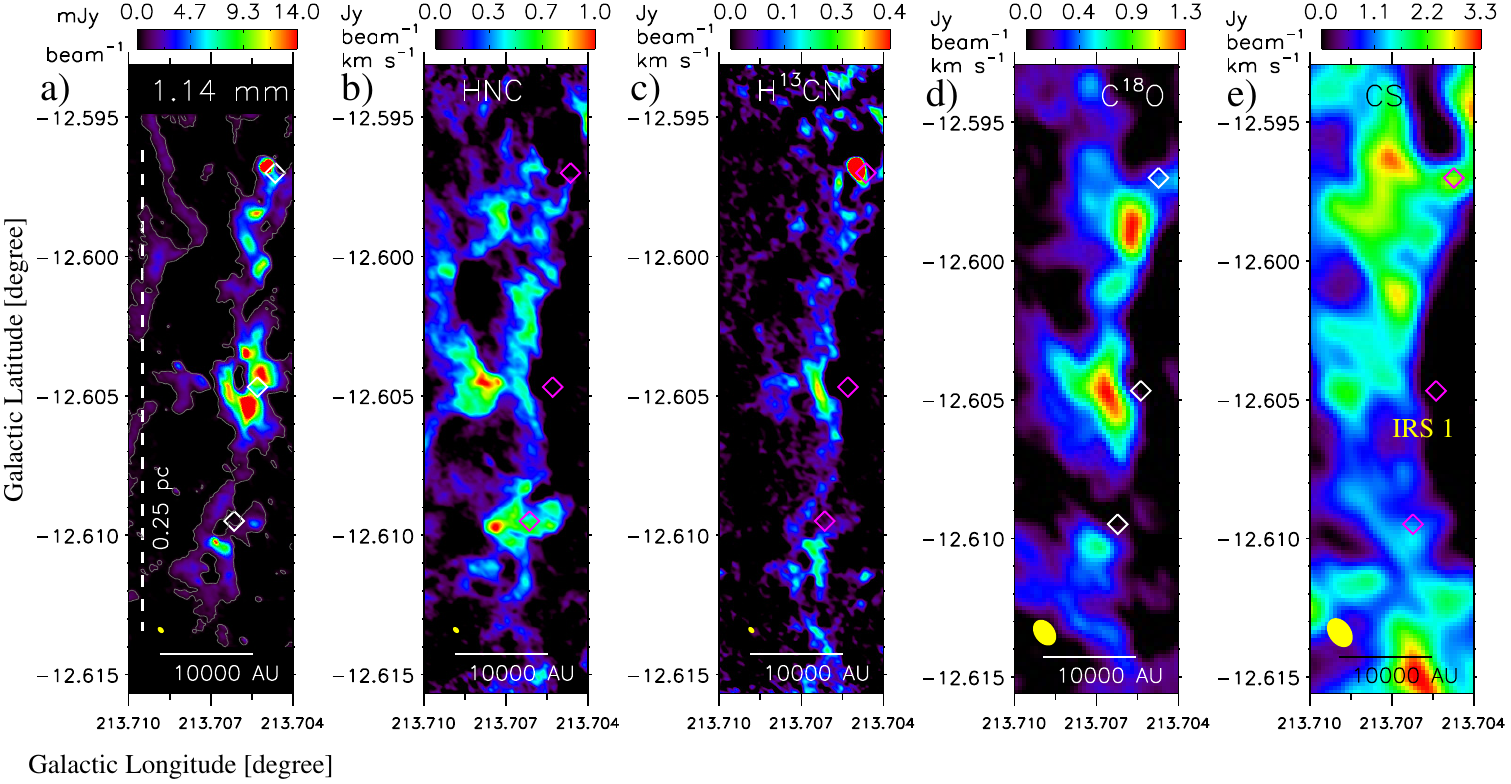}
\includegraphics[width=12 cm]{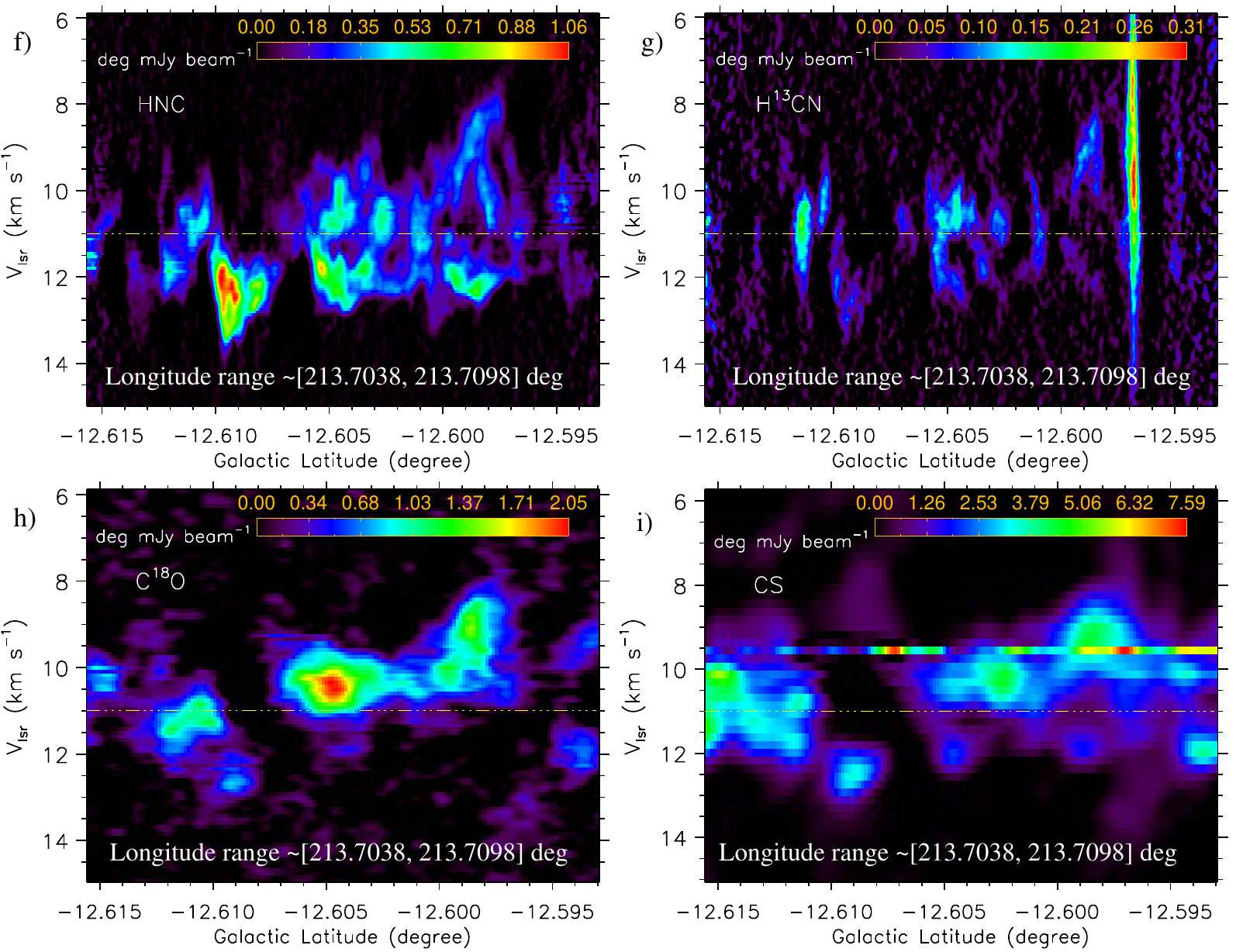}
\caption{Distribution of dust and molecular emissions in the direction of an area highlighted by the solid box in Figure~\ref{fg1}d.
a) The panel shows the ALMA 1.14 mm dust continuum map and the 1.14 mm continuum emission contour at 0.45 mJy beam$^{-1}$. 
Integrated intensity map (at $V_{\rm lsr}$ of [6, 15] km s$^{-1}$) of b) HNC(3--2); 
c) H$^{13}$CN(3--2); d) C$^{18}$O(1--0); e) CS(2--1). 
In panels ``a--e'', diamonds are the same as in Figure~\ref{fg1}a. 
Latitude--velocity maps of f) HNC(3--2); g) H$^{13}$CN(3--2); h) C$^{18}$O(1--0); i) CS(2--1).
In panels ``f--i'', the molecular emission (at $V_{\rm lsr}$ of [6, 15] km s$^{-1}$) is integrated over the
longitude from 213.7038 deg to 213.7098 deg (see the solid box in Figure~\ref{fg1}d).
}
\label{fg10}
\end{figure*}
\begin{figure}
\center
\includegraphics[width=8cm]{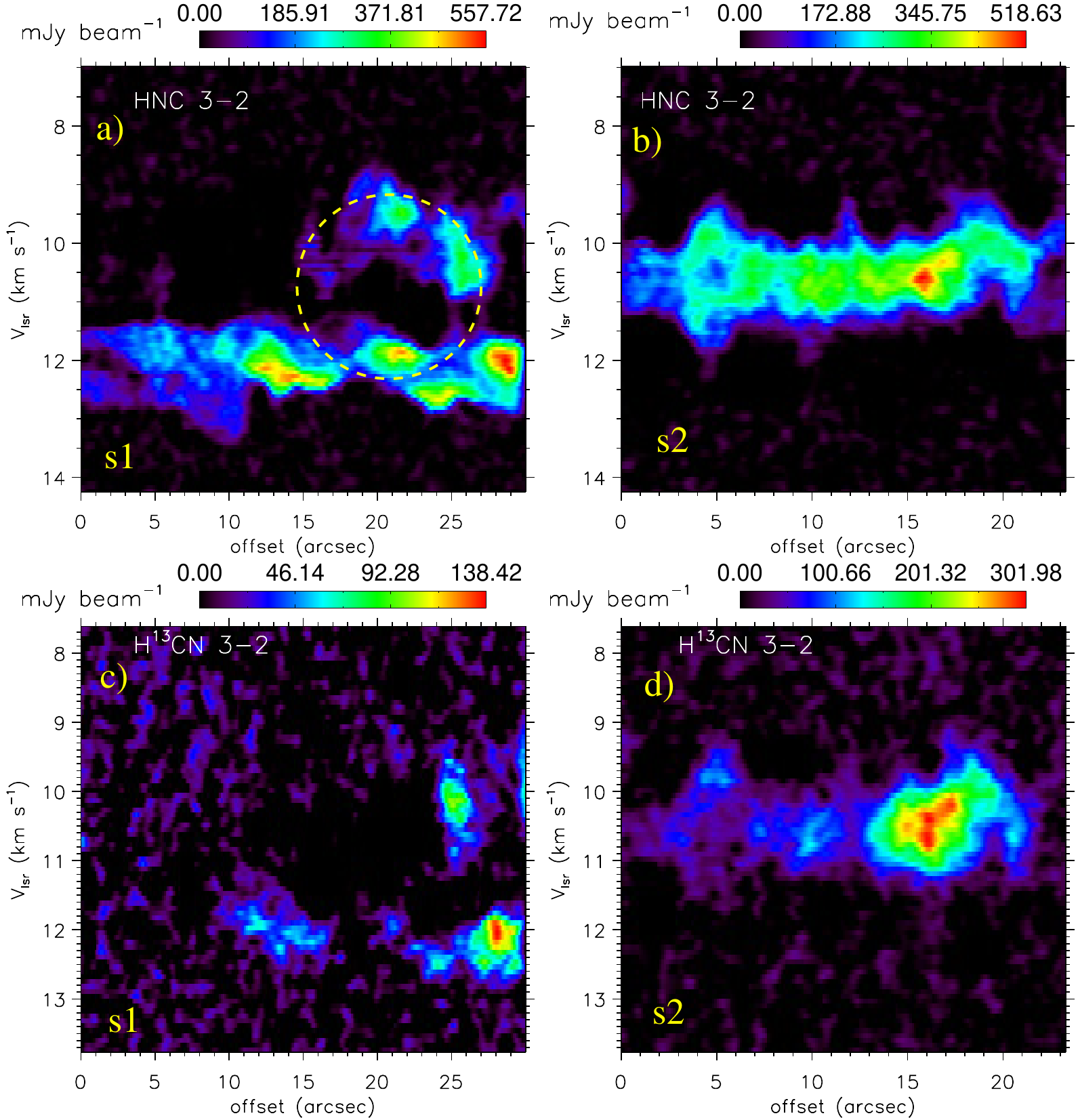}
\includegraphics[width=8cm]{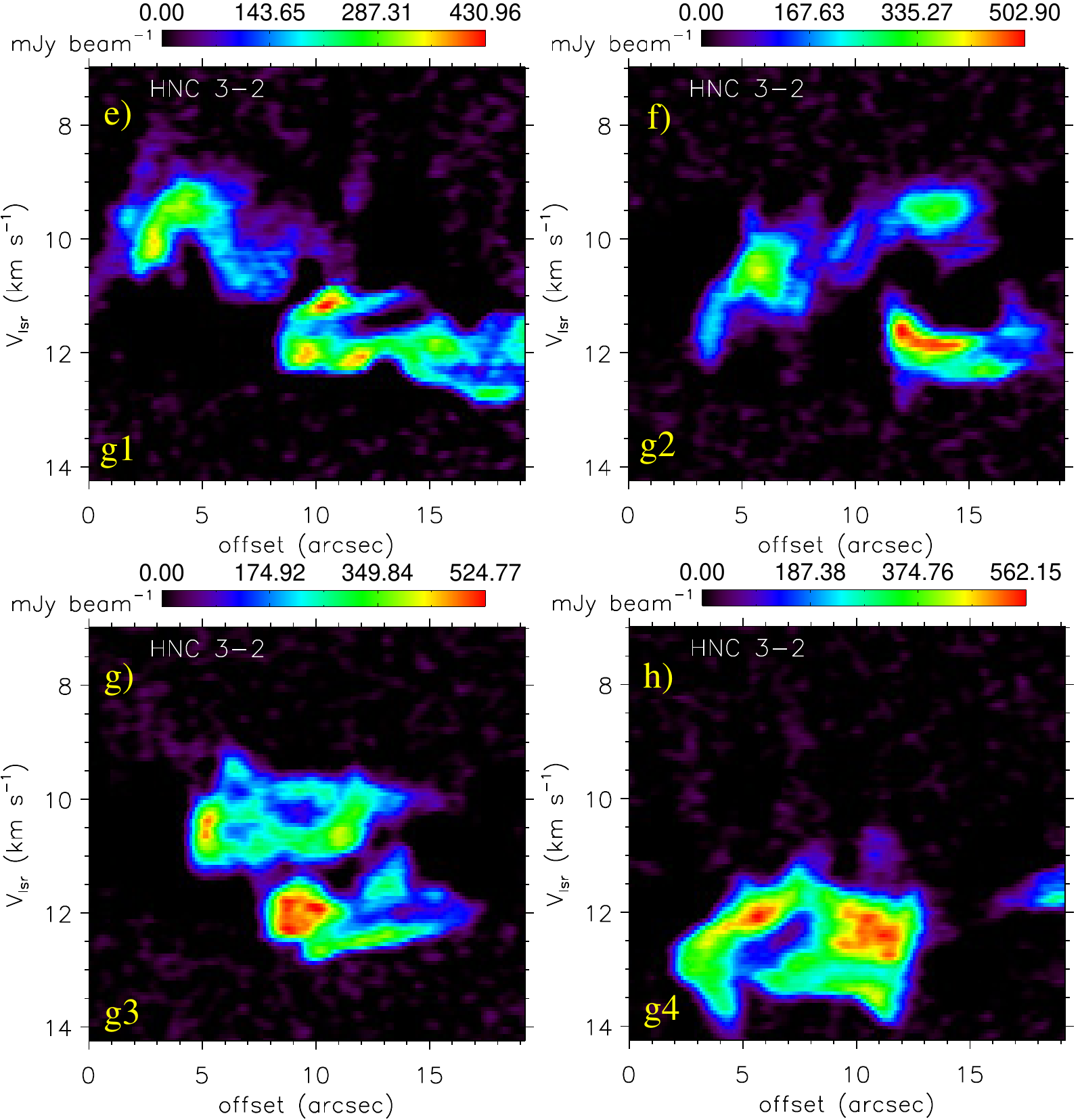}
\caption{PV diagram of the a) HNC(3--2); c)H$^{13}$CN(3--2) emission along the path ``s1''. 
PV diagram of the b) HNC(3--2); d)H$^{13}$CN(3--2) emission along the path ``s2''. e--h) PV diagrams of the HNC(3--2) along the paths``g1--g4''. All these paths are indicated by dot-dashed lines in Figure~\ref{fg5}a.} 
\label{fg8}
\end{figure}
%
%
\subsubsection{Channel maps and PV daigrams}
\label{zsubsec:data3j} 
To probe the gas distribution toward the mirrored B-like feature, 
%
in Figure~\ref{fg7}d, we display the ALMA HNC(3--2) filled contour map at 10.68 km s$^{-1}$ (in green) overlaid with contours of HNC(3--2) emission at 12.16 km s$^{-1}$ (in red) and 9.94 km s$^{-1}$ (in blue). These velocity channels are selected based on the different peaks seen in the average HNC(3--2) emission profile (see Figure~\ref{fg7}e). 
The average profile is extracted in the direction of an area highlighted by the rectangle box in Figure~\ref{fg7}d, which is selected toward the central part of the mirrored B-like feature. From Figure~\ref{fg7}d, we infer that the outer region of the mirrored B-like feature 
is associated with gas at 12.16 km s$^{-1}$, while gas at 9.94 and 10.68 km s$^{-1}$ is traced toward its base segment. Interestingly, the central part of the mirrored B-like feature (or the dust continuum source ``r1'') appears to be associated with gas at different velocities. 

To conduct a comparative study, we present the 1.14 mm dust continuum map, moment-0 maps (at $V_{\rm lsr}$ of [6, 15] km s$^{-1}$) of HNC(3--2), H$^{13}$CN(3--2), C$^{18}$O(1--0) and CS(2--1) in Figures~\ref{fg10}a,~\ref{fg10}b,~\ref{fg10}c,~\ref{fg10}d, and~\ref{fg10}e, respctively. Note that the selected area, which is indicated by the solid box in Figure~\ref{fg1}d, emcompasses the dust filament and the mirrored B-like feature. The C$^{18}$O(1--0) and CS(2--1) emission maps seem to depict only the base segment and central part of the mirrored B-like feature. In Figures~\ref{fg10}f,~\ref{fg10}g,~\ref{fg10}h, and~\ref{fg10}i, we show the latitude--velocity maps of the HNC(3--2), H$^{13}$CN(3--2), C$^{18}$O(1--0) and CS(2--1) emission, respectively. 
The diagrams of the C$^{18}$O(1--0) and CS(2--1) emission are shown here only for a comparison purpose. 
A semi circular velocity structure is found toward the latitude of [$-$12.603, $-$12.606] degrees in the HNC(3--2) and H$^{13}$CN(3--2) emission, which covers the central part of the mirrored B-like feature. The outflow activity is clearly seen toward IRS~3 in the PV diagram of the H$^{13}$CN(3--2) emission. 
In the direction of IRS~4, a noticeable velocity gradient is also seen in the velocity space.

To examine PV diagrams of the HNC(3--2) emission toward 
the mirrored B-like feature, several paths (s1, s2, g1, g2, g3, and g4; see dot-dashed lines in Figure~\ref{fg5}a) are arbitrarily chosen. 
Figures~\ref{fg8}a and~\ref{fg8}b display PV diagrams of the HNC(3--2) emission along the paths ``s1'' and  ``s2'', respectively. In Figures~\ref{fg8}c and~\ref{fg8}d, we present PV diagrams of the H$^{13}$CN(3--2) emission along the paths ``s1'' and  ``s2'', respectively. The path ``s1'' starts from the outer region of the mirrored B-like feature and ends at its central part, while the path ``s2'' passes toward the base segment of the mirrored B-like feature. From Figure~\ref{fg8}a, we find a velocity structure around 12 km s$^{-1}$, characterized by two sub-features crossing each other, resembling a braid. 
Furthermore, in the direction of the central part of the mirrored B-like feature, an almost circular velocity structure is seen in Figure~\ref{fg8}a (see the dashed circle). 
Figure~\ref{fg8}b displays a velocity structure around 10.5 km s$^{-1}$, also with two sub-features intersecting like a braid. These velocity structures are absent in the PV diagrams of the H$^{13}$CN(3--2) emission.
We have also generated PV diagrams along 4 paths (see ``g1--g4'' in Figure~\ref{fg5}a), which are perpendicular to the mirrored B-like feature. These diagrams along the paths``g1--g4'' are shown in Figures~\ref{fg8}e--\ref{fg8}h, and show the velocity features around 10 and 12 km s$^{-1}$. 
Each velocity feature observed toward the paths g1, g2, and g3 has two sub-features that intertwine like a braid. 
\subsubsection{Position-position-velocity maps}
\label{zsubsec:data3q} 
\begin{figure*}
\center
\includegraphics[width=\textwidth]{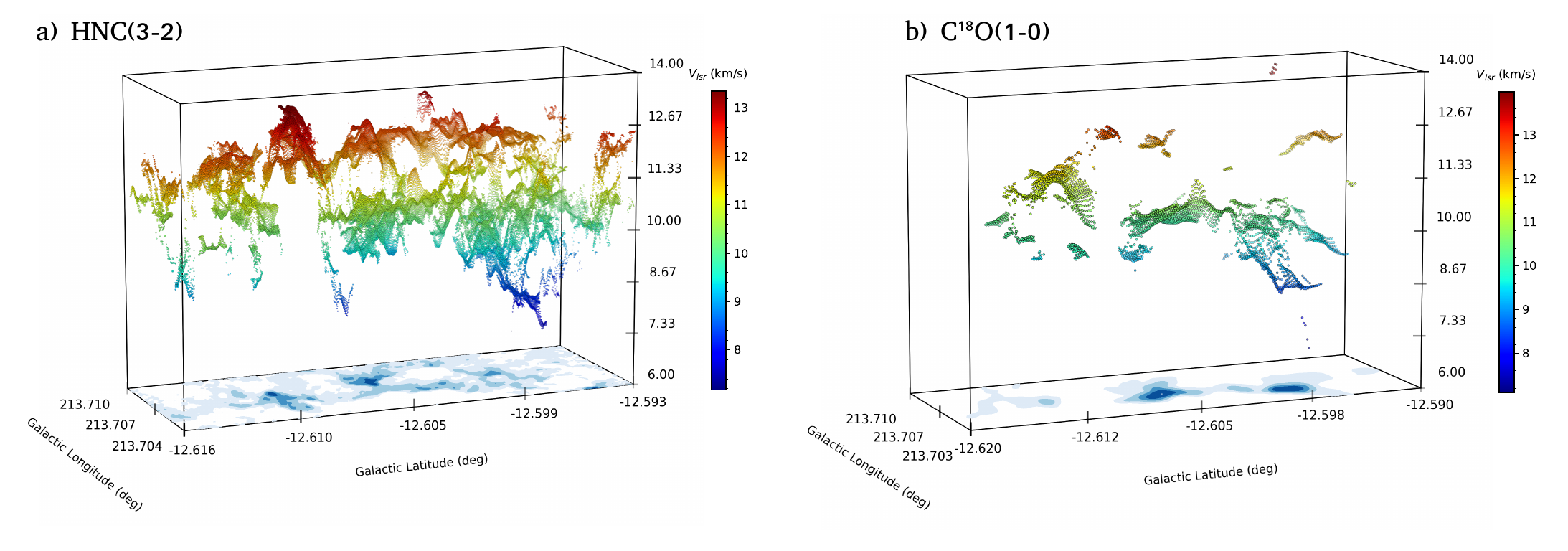}
\caption{PPV maps of a) ALMA HNC(3--2) and b) ALMA C$^{18}$O(1-0) in the direction of an area shown in Figures~\ref{fg10}b and~\ref{fg10}d. 
The PPV maps are produced using the tool \texttt{SCOUSEPY}. }
\label{fg11}
\end{figure*}

Figures~\ref{fg11}a and~\ref{fg11}b display position-position-velocity (PPV) maps of the HNC(3--2) and C$^{18}$O(1--0) emission in the direction of an area shown in Figure~\ref{fg10}b, respectively. These maps are produced using the tool \texttt{SCOUSEPY}  \citep{henshaw16,henshaw19,dewangan24}, which is used to perform the spectral decomposition of the complex spectra. In this analysis, we start the decomposition of the spectra by defining the size of the `Spectral Averaging Areas (SAAs)' in pixels, which are 5 $\times$ 5 pixel$^2$ for C$^{18}$O(1--0) and 15 $\times$ 15 pixel$^2$ for HNC(3--2). These SAAs are arranged to cover all emissions above a user-defined emission level (i.e., 0.2~Jy beam$^{-1}$ for C$^{18}$O(1--0) and 0.13~Jy beam$^{-1}$ for HNC(3--2)). This threshold is selected to ensure that the morphology in the position--position (i.e., {\it l--b}) plane closely aligns with that of the moment-0 map. An averaged spectrum is extracted from the SAAs and fitted with single or multiple Gaussian components (if present). Thereafter, all the pixels within each SAA are fitted with multiple velocity components for their respective SAA. To better understand the resulting output, we plot the centroid velocity of the fitted Gaussian(s) in each pixel in the PPV space. We have also shown the moment-0 map with emission above 3$\sigma$ in the {\it l--b} plane at the bottom of the PPV map.

%
%
The gas at [10, 11] km s$^{-1}$ is detected in both the HNC(3--2) and C$^{18}$O(1--0) emissions which trace the vertical structure on the right (i.e., the base of the mirrored B-like feature), as evident from Figure~\ref{fg7}d. On the other hand, the gas around 12 km s$^{-1}$ is abundant in HNC(3--2) but nearly absent in the C$^{18}$O(1--0) emission. The same gas component at approximately 12 km s$^{-1}$ also emits in CCH (4--3) and (3--2), as shown in Figure~\ref{fg5}. All these emission lines are tracers of PDRs, indicating that different physical conditions and chemistry are present in the vertical structure of the B-like feature on the right compared to the left lobes. A noticeable velocity variation is seen toward the gas around 12 km s$^{-1}$. 
In the direction of the base of the mirrored B-like feature, we find the velocity oscillations between 10 and 11 km s$^{-1}$ in the HNC(3--2) emission (see Figure~\ref{fg11}a). In the direction of the mirrored B-like feature, the sub-structures detected in PV diagrams appear to be linked with the velocity oscillations observed in PPV maps.  
\subsection{Integrated multi-scale view of Mon~R2 }
\label{subsec:discxc1}
\begin{figure}
\center
\includegraphics[width=8 cm]{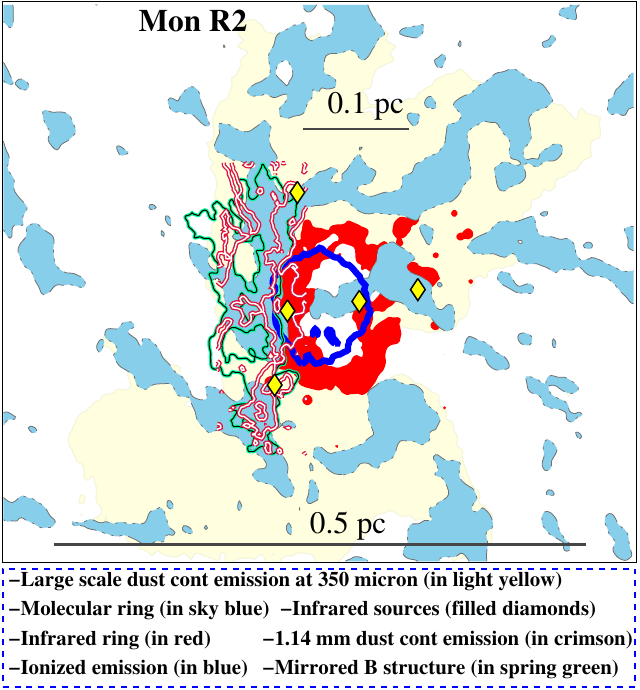}
\caption{This panel summarizes the various features observed toward the Mon~R2 in this study.}
\label{fg11c}
\end{figure}

Taking into account the derived outcomes at different physical scales, a summary figure is presented in Figure~\ref{fg11c}, 
where the locations of different observed structures toward Mon R2 are highlighted. 
It includes the large scale dust continuum structure, molecular ring, IR ring, mirrored B-like feature, IR sources, dust filament, ionized spherical shell, and small-scale dust shell-like feature. 

In Mon~R2, both the molecular ring (size $\sim$0.18 pc $\times$ 0.26 pc) and IR ring (size $\sim$0.12 pc $\times$ 0.16 pc) show an elliptical morphology, though they differ in size (see Figure~\ref{fg11c}). The dense gas tracer, C$^{18}$O line emission, is detected in the central part of the molecular ring, which also corresponds to the central area of the IR ring. In the direction of the IR ring, the ionized shell appears spherical with an extent of $\sim$0.12 pc, and the small-scale dust shell-like feature (extent $\sim$0.04 pc $\times$ 0.07 pc) hosting IRS~1 is seen. The implications of these observed features can be found Section~\ref{subsec:disc1sub}.

\begin{figure}
\center
\includegraphics[width=8cm]{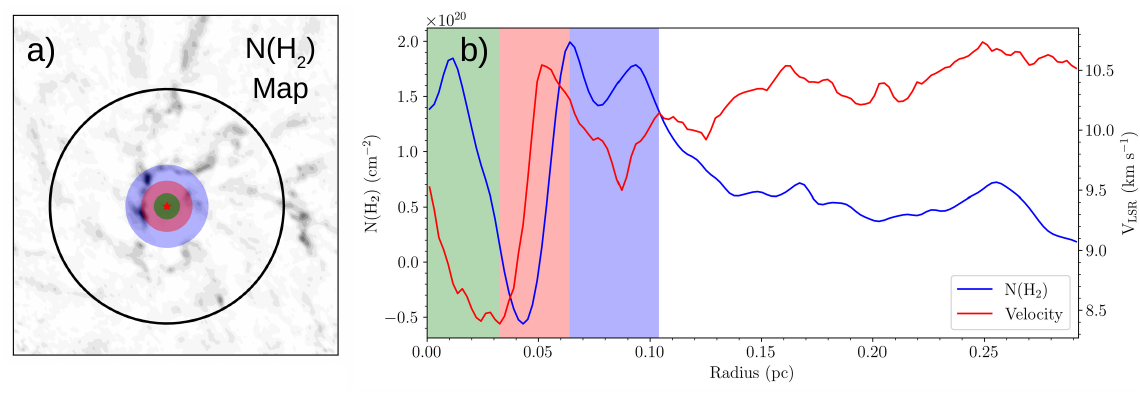}
\includegraphics[width=8cm]{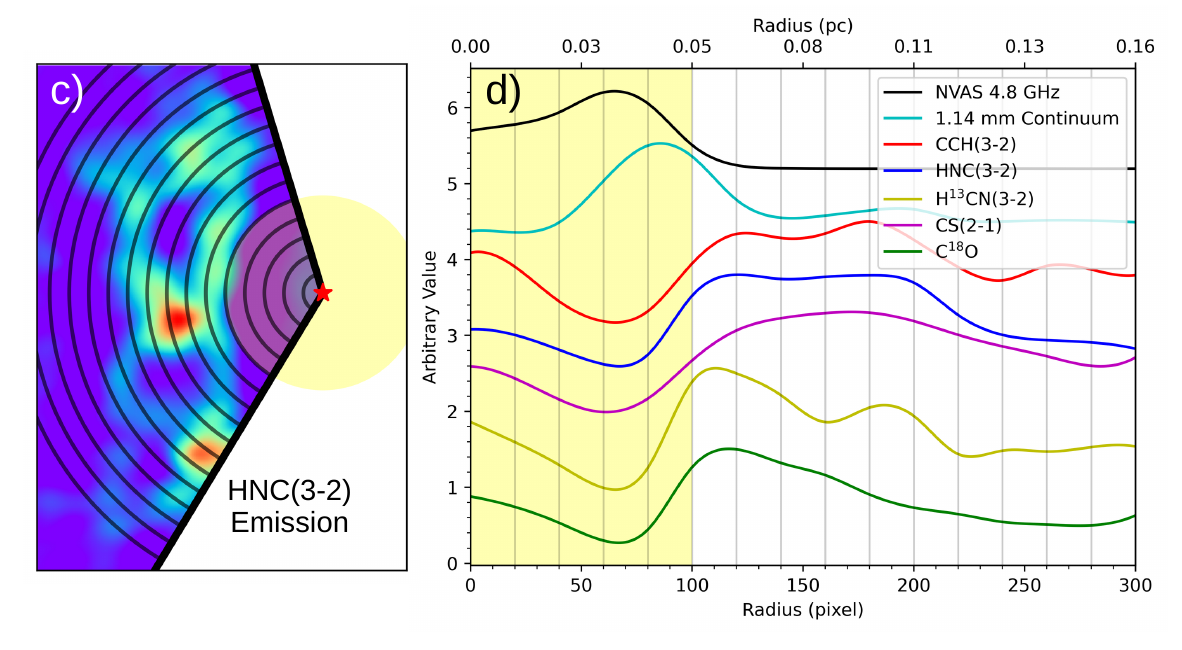}
\caption{a) The column density ($N$(H$_2$)) gray scale map derived using C$^{18}$O(1--0). 
Different zones are highlighted by open and filled circles centered on IRS 2 where the radial distribution of $N$(H$_2$) and $V_{\rm lsr}$ is examined (see Figure~\ref{fg12}b). b) Radial distribution of $N$(H$_2$) and $V_{\rm lsr}$ estimated from the circle center marked by a star indicated in Figure~\ref{fg12}a. 
Colored regions in the radial plot refer to the areas marked in Figure~\ref{fg12}a. c) The panel displays the integrated intensity map of HNC(3--2) overlaid with several concentric circular arcs centered on IRS~2 (star), where the radial distribution of different emission toward the hub region in Mon R2 is examined (see Figure~\ref{fg12}d). 
d) Radial distribution of various emissions toward the hub region in Mon R2 estimated toward the eastern zone from the circle center marked by a star indicated in Figure~\ref{fg12}c. Concentric circular arcs highlighted in Figure~\ref{fg12}c are marked by vertical lines in the radial plot. In panels ``a'' and ``c'', the maps are the same resolution of about 4$''$.} 
\label{fg12}
\end{figure}
\subsection{Radial distribution of velocity and column density in Mon R2}
\label{tsubsec:datad} 
To study the gas properties in the vicinity of Mon R2 hub, we selected a circular area centered at the position of IRS~2 (see the star in Figure~\ref{fg12}a). It is important to note that the position of IRS~2 roughly corresponds to the
geometrical center of the cluster in Mon R2 \citep[see also][and references therein]{carpenter08}.  
We examined the radial distribution of $N\rm (H_{2})$ and $V_{\rm lsr}$ of the C$^{18}$O emission with respect to this position. The C$^{18}$O moment-1 map is used for the $V_{\rm lsr}$ distribution. Figure~\ref{fg12}a presents the $N\rm (H_{2})$ map overlaid with a circular boundary of $\sim$0.3 pc radius, where the radial distributions of $N\rm (H_{2})$ and $V_{\rm lsr}$ are examined and shown in Figure~\ref{fg12}b. The radial profiles represent azimuthally averaged values calculated for each one-pixel wide circular annulus, extending up to $\sim$0.3 pc radius. Based on distinct features observed in both the radial profiles and the spatial distribution of gas, we have highlighted three colored circular annuli (or zones) in Figures~\ref{fg12}a and~\ref{fg12}b. The green zone traces the immediate vicinity of IRS~2. 
The gas blob associated with IRS~2 is detected at a $V_{\rm lsr}$ of around 9 km s$^{-1}$, while the red zone is primarily gas deficit and shows a transition to gas/material with $V_{\rm lsr}$ of around 10 km s$^{-1}$. In this region, the distributions of $V_{\rm lsr}$ and $N\rm (H_{2})$ show similar trends but keep a lag which might be indicative of continuous gas expansion. 
The blue zone encloses the filament/ridge along with the sources IRS~1 and IRS~3, where the gas material is observed at a $V_{\rm lsr}$ of around 10 km s$^{-1}$, with variations of around 1 km s$^{-1}$. 
For the outer region, the radial distributions of $V_{\rm lsr}$ and $N\rm (H_{2})$ display anticorrelation, which is one of the important results of this paper (see Section~\ref{subsec:spiral} for more details).  
\section{Discussion}
\label{sec:disc}
In the literature, it has been reported that  in Mon~R2 HFS, the young UCH\,{\sc ii} region \citep[age $\sim$10$^{5}$ yr;][]{didelon15} has begun to expand and break out of the dense filamentary hub where it formed, while material continues to collapse inward along the filaments \citep[e.g.,][]{pilleri14,morales16,morales19}. Consequently, the Mon R2 HFS represents a system where the interaction between gas accretion through filaments and feedback from massive stars is evident. 
This work offers new insights into various observed structures for the first time and their implications for understanding the ongoing physical processes in Mon R2. 
\subsection{Spiral-like structure in Mon R2}
\label{subsec:spiral}
Both the SHARC-II continuum map at 350 $\mu$m and the ALMA C$^{18}$O emission support the presence of earlier reported spiral-like structure or Mon R2 HFS (see Figure~\ref{fg1}d). The molecular ring is interconnected by molecular or dust filaments (or hub-composing filaments) in every direction, outlining the physical extent of the central hub in the Mon~R2 HFS (see Figure~\ref{fg11c}). 
The spiral structure could be a result of the rotation of the cloud or clump at a 1 pc scale \citep{morales19,hwang22} and may be related to the formation of HFSs \citep[see][and references therein]{maity24}.
From Figure~\ref{fg1}d, it is clearly seen that the central area of the spiral-like structure lacks a tightly compact structure, which may be attributed to a combination of factors such as mass distribution, thermal pressure, turbulence, magnetic fields, stellar feedback, and evolutionary stage of the system. 

In any HFS, assessing the distribution of material between the central hubs and the filaments is essential \citep[e.g.,][]{morales19,kumar22}. 
An examination of the SHARC-II continuum map at 350 $\mu$m reveals that the center of the spiral structure contains more mass than its filaments (see Section~\ref{s3sec:d}). 
This mass distribution might support efficient transport mechanisms that channel material along the filaments toward the center, where the presence of low-mass and high-mass stars is evident. According to one statistical study of HFSs, multi-scale mass accretion/transfer flows, i.e. accretion from clumps onto cores and that from cores to embedded protostars, are the major driver for massive star formation in HFSs \citep{liu23}. 

The ALMA C$^{18}$O and CS line data show the variations of velocity toward the molecular ring. 
In the direction of one of the junctions (see the dotted curve in Figure~\ref{fg3}a), a V-shaped velocity feature is seen in PV diagrams of C$^{18}$O and CS (see Figures~\ref{fg3}c and~\ref{fg3}d). 
Such features hint on the mass accretion toward the molecular ring along the molecular filament \citep{zhou23}.
In this relation, the observed velocity variations in the C$^{18}$O and CS emissions may hint the accretion flow signatures (see also Figures~\ref{fg12}a and~\ref{fg12}b). 
In Section~\ref{tsubsec:datad}, an anticorrelation is also observed between the radial distributions of $N\rm (H_{2})$ and $V_{\rm lsr}$, derived from the dense gas tracer (see Figure~\ref{fg12}b). 
It hints that low column-density filaments are channeling inflowing material into the central hub-region. 

\subsection{Distinct Ring-like features in Mon~R2 }
\label{subsec:disc1sub}
In the Mon R2 HFS, the molecular ring encompasses the IR ring, and a separation has been observed between these two rings (see Section~\ref{subsec:discxc1}). 
This outcome contrasts sharply with the commonly reported findings for mid-IR (MIR) bubbles, which coexist with molecular gas \citep{churchwell06,churchwell07}. 
This provides a direction for further studies to distinguish bubbles formed within the hub from those in typical molecular clouds.
%

The MIR bubbles typically appear as ring-like or shell-like structures in the {\it Spitzer} 8 $\mu$m images, and often surround H\,{\sc ii} regions (or massive OB stars). The bubbles are thought to result from the feedback of massive OB-type stars that push away surrounding materials \citep{deharveng10}. The edges of these bubbles 
are often associated with dense molecular materials and show signs of ongoing star formation. These signatures may correspond to a ``collect and collapse'' scenario, where gravitational instabilities lead to the fragmentation of the accumulated ring into molecular condensations, which then further fragment into the cores observed \citep[e.g.,][]{deharveng03}. Most recently, hierarchical triggering was reported in a ring-like H\,{\sc ii} region G24.47+0.49 \citep{saha24}. 
Taking into account the hub-filament configuration and the observed gap between the molecular ring and the IR ring, the ``collect and collapse'' scenario may not be applicable in Mon~R2. 

In the ionized spherical morphology, radio continuum emission is mostly found at opposite edges containing IRS~1 and IRS~2 rather than at the center. The projected separation of the two massive stars IRS1 and IRS2 is $\sim$0.07 pc. 
Each edge displays arc-like radio emission, centered around one IRS source, facing the central region of the ionized spherical morphology (see the inset in Figure~\ref{fg1}a). This could hint at the H\,{\sc ii} region's expansion and the counterbalancing effects exerted by the surrounding dense molecular environment (i.e., molecular ring). 

The existence of the extended PDR in Mon R2 has been proposed \citep[see][and references therein]{ginard12} and its projected thickness ranges between 4$''$ and 6$''$ \citep[e.g.,][]{berne09,pilleri14}. It implies the presence of the molecular/ionized gas interface where the influence of UV radiation from massive stars is expected. 
In Appendix~\ref{s4sec:d}, our analysis related to various pressure components also suggests that the existence of the IR ring in Mon~R2 can be attributed to the feedback from massive stars, primarily IRS 1 and IRS 2. 
Additionally, the dense molecular ring also appears to be affected by the impact of massive stars (see Appendix~\ref{s4sec:d} for more details). 

In the direction of the mirrored B structure, blueshifted and redshifted gas components in the HNC(3--2) emission are observed (see Figure~\ref{fg7}d). The PV diagram of HNC(3--2) reveals circular velocity features. 
These findings, based on ALMA band-6 molecular line data, provide evidence for an expanding H\,{\sc ii} region in Mon R2 (see Section~\ref{subsec:disc4r} for more discussion). 
\subsubsection{Eastern part of molecular ring: mirrored B structure}
\label{subsec:disc4r}
Using the HNC(3--2) line data, we have investigated the mirrored B-like feature (extent $\sim$19000 AU $\times$ 39000 AU) toward the eastern part of the molecular ring (see Section~\ref{zsubsec:data3i}). The HNC(3--2) line data are often used as PDR tracer, and hence can enable us to depict the distribution and interaction of cold, dense gas with UV radiation. According to \citet{berne09}, in Mon~R2, the interface between the ionization front and the dense molecular gas is marked by a dense PDR, characterized by parameters such as $n$ = 4 $\times$ 10$^{5}$ cm$^{-3}$, $N$(H$_2$) = 1 $\times$ 10$^{21}$ cm$^{-2}$ and $T_{rot}$ = 574($\pm$20)~K. This region is well-detected in pure rotational H$_{2}$ lines, MIR polycyclic aromatic hydrocarbon (PAH) bands, and rotational lines of reactive molecular ions like CO$^{+}$ and HOC$^{+}$ \citep[e.g.,][]{berne09}. 

The moment maps, PV diagram, and PPV maps of the HNC(3--2) emission show the presence of 
blueshifted gas toward the base segment of the mirrored B structure and redshifted gas toward the outer part of the mirrored B structure (see Section~\ref{subsec:data3b}). 
It implies an expansion of the molecular gas toward the eastern part of the molecular ring.
Furthermore, the PV diagram reveals an almost circular velocity structure toward the latitude of [$-$12.603, $-$12.606] degrees in the HNC(3--2) and H$^{13}$CN(3--2) emission, which covers the central part of the mirrored B-like feature. 
This can be explained with the results of modelling of expanding shells as discussed in \citet{arce11} \citep[see also][]{dewangan16,saha24}. 
On the basis of the maximum redshifted and blueshifted velocity components in the PV diagram, we compute an expansion velocity of $\sim$2.25 km s$^{-1}$ in the direction of the eastern part of the molecular ring. 

In the base segment of the mirrored B structure, IRS 3, IRS 4, and multiple condensations are evident (see Figure~\ref{fg4}c and also Table~\ref{tab2}).  
Additionally, the half part of the dust shell-like feature, containing the compact dust continuum source ``c4'' without radio continuum emission, lies almost at the center of the base segment. Approximately the other half of this structure hosting the compact dust continuum sources (i.e., ``c1--c3'' and ``c5'') shows no connection with molecular emission, but it does exhibit radio continuum emission. Figure~\ref{fg12}c presents the overlay of concentric circular arcs (20-pixel width) on the HNC(3--2) emission map for which the radial distribution of different emissions is examined with respect to the position of IRS~2. To plot the radial profiles, we have smoothed all the maps to the resolution of C$^{18}$O emission and re-gridded over the pixel scale of the HNC map. We further smoothed the radial curves to get rid of sharp intensity variations.
Figure~\ref{fg12}d shows the radial distribution of ionized gas (NVAS 4.8 GHz), dust emission (1.14 mm continuum), PDR molecules (CCH(3--2), HNC(3--2)), and dense molecular gas (H$^{13}$CN, CS, C$^{18}$O). The boundary of the yellow circle shown in Figure~\ref{fg12}c is also marked in Figure~\ref{fg12}d. 
The radio peak emission is observed within the circular area, followed by all subsequent molecular emissions.
Dust peak emission is observed both just inside and outside the edge of the circular area, with its peak offset from the radio peak emission. 
In other words, the dust emission appears to be situated between the ionized and molecular emissions. 

From Figure~\ref{fg12}d, it is evident that PDRs predominantly extend beyond this circular region or ionized area, 
where we find the existence of sub-structures in both redshifted and blueshfited velocity components. In each velocity component, we find these sub-features intersecting like a braid. This argument is further supported with the observed velocity oscillations toward the base segment and the outer part of the mirrored B structure (see Section~\ref{zsubsec:data3q}). One might generally expect an oscillation-like velocity pattern, potentially associated with the twisted 
behavior \citep[i.e.,][]{dewangan21}. 
The study of H\,{\sc ii} regions powered by massive OB stars generally provides an opportunity to investigate the transitional boundary between the neutral/molecular PDR and the fully ionized H\,{\sc ii} region. 
Variations in temperature and density among the gas layers in the PDR can cause instabilities, leading to the development of complex structures \citep[e.g.,][]{goicoechea16,carlsten18,wolfire22,Dewangan2023_NGC_3324}.
Using the NIR and MIR images from the James Webb Space Telescope (JWST), \citet{Dewangan2023_NGC_3324} examined the PDR of the massive star-forming region NGC 3324 in the Carina Nebula. They investigated intertwined/entangled sub-structures in H$_{2}$ toward the bubble wall of NGC 3324 at scales below 4500 AU. These sub-structures in the PDRs could be treated as direct indicators of instability at the dissociation front. Hence, in this work, the observed velocity sub-structures support the presence of instability in the PDR. 

In the UCH\,{\sc ii} Mon R2 region, \citet{morales16} studied the layer between the H\,{\sc ii} region and the molecular gas using the CO$^{+}$ map, 
which shows a clumpy ring-like distribution that is spatially coincident with the PAHs emission at 11.3 $\mu$m (see Figure~1 in their paper). 
Different physical conditions are present in the extended PDR of UCH\,{\sc ii} region powered by IRS 1 \citep[e.g.,][]{ginard12,pilleri12,pilleri13,pilleri14}. 
\citet{pilleri13} presented a 1D schematic of the geometry of Mon R2 as a 
function of the radial distance to IRS 1 \citep[see Figure~2 in their paper and also][]{pilleri12}. 
According to these authors, this PDR may either represent the outer layer of the molecular cloud exposed to an external UV field or be the result of UV photons escaping through a gap near IRS 1 and illuminating the surrounding walls. 
Our findings seem to support the proposal of \citet{pilleri12,pilleri13} and the PDR in Mon R2 corresponds to the ionized spherical shell that hosts IRS 1 and IRS 2.
\subsection{Is the Mon R2 HFS an example of an IR-quiet to IR-bright HFS ?}
\label{subsec:disc2a}
It is believed that HFSs play a key role in our understanding of how massive stars and groups of stars are formed \citep[][]{kumar20,zhou22,liu23}. Different scenarios to explain the HFSs are presented in Section~\ref{sec:intro}. 
In the GNIC scenario, containing the flavours of the CA and GHC models, \citet{Tige+2017} and \citet{Motte+2018} described that in a hub/ridge filament system \citep[see also][]{morales19}, gas flows along the filaments into the central hub, where massive dense cores (MDCs) form at scale $\sim$0.1 pc. Initially starless for $\sim$10$^{4}$ years, MDCs evolve into protostellar phases once a low-mass stellar embryo forms, lasting $\sim$3 $\times$ 10$^{5}$ years. When the embryo grows beyond 8 M$_{\odot}$, it becomes IR-bright followed by creating an H\,{\sc ii} region at later stages.

Examining the presence of extended H\,{\sc ii} regions around high-luminosity sources in central hubs helps study the evolutionary stages of HFSs. The IR-quiet HFS lacks such sources, while the IR-bright HFS contains them, leading to ionization of their surroundings. Early-stage hubs are compact and dense, with minimal ionization (i.e., IR-quiet HFS), while later stages are more extended, shaped by feedback processes and expanding ionized regions (i.e., IR-bright HFS). This approach provides insight into the evolution of HFSs and the role of massive star feedback in shaping the HFSs.

On a spatial scale of under 0.6 pc, most recently, using high-resolution NIR images from the JWST, \citet{dewangan24} discovered an IR-dark HFS candidate (extent $\sim$0.55 pc; referred to as G11P1-HFS) toward a promising massive protostar G11P1 in G11.11$-$0.12. The center of G11P1-HFS hosts embedded NIR sources associated with radio continuum emission, displaying the presence of forming massive stars in a dense, compact hub with minimal ionization. 
In contrast, on a spatial scale of less than 0.8 pc, Mon R2 HFS can now be classified as an IR-bright HFS. 
This HFS contains high-luminosity sources along with low-mass stellar populations, with the impact of massive stars clearly observed. When comparing these two HFSs, Mon R2 HFS seems to be more evolved than G11P1-HFS.

%
%
%
Our analysis of various line and continuum data reveals that Mon R2 HFS was once an IR-quiet HFS, which has now evolved into an IR-bright HFS. This is evidenced by dark areas in NIR images, showing dense molecular gas and low-mass objects with IR excess, especially at the center of the molecular and IR rings. The molecular ring marks the hub boundary, where IR excess is observed near IRS 2, while IRS 1 is embedded in a small-scale dust shell. Both sources are associated with radio continuum emission. 
Massive stars within the shell leads to the creation of an H\,{\sc ii} region, with their feedback responsible for the IR ring and the expansion of the hub. PDRs and instability signatures are seen at the molecular ring's edge, possibly pushed outward by feedback as the H\,{\sc ii} region expanded. Despite these changes, accretion into the system continues, illustrating the dynamic interplay between gas accretion and massive star feedback in Mon R2 HFS.
\section{Summary and Conclusions}
\label{sec:conc}
To obtain a comprehensive understanding of the physical processes occurring in previously known Mon R2 HFS, a multi-wavelength and multi-scale study has 
been conducted. This study integrates data sets from several facilities/surveys including ALMA, NVAS, {\it HST}, SHARC-II, and UKIDSS. 
These diverse data sets have enabled a thorough examination of the distribution of dust, molecular gas, ionized emission, and embedded stellar populations toward the Mon~R2 HFS. The key findings of this research are outlined as follows:\\
$\bullet$ The SHARC-II dust continuum map at 350 $\mu$m and the ALMA C$^{18}$O emission map at [8, 13] km s$^{-1}$ reveal a spiral-like structure of Mon R2 HFS. 
The center of the spiral structure contains more mass than its filaments.\\
$\bullet$ The ALMA C$^{18}$O(1--0) emission map depicts a molecular ring  (size $\sim$0.18 pc $\times$ 0.26 pc) interconnected by filaments, marking the hub's extent. The hub in the Mon~R2 HFS does not exhibit a tightly compact structure.\\
$\bullet$ Using the ALMA C$^{18}$O line data, a V-shaped velocity feature and an anticorrelation between the radial distributions of $N\rm (H_{2})$ and $V_{\rm lsr}$ are investigated,  suggesting mass accretion toward the molecular ring along the molecular filaments. \\
$\bullet$ The IR ring (size $\sim$0.12 pc $\times$ 0.16 pc) and the spherical ionized morphology (extent $\sim$0.12 pc) with radio emission concentrated at the edges are both enclosed by the molecular ring. This may suggest the expansion of the H\,{\sc ii} region.\\
$\bullet$ The IR ring surrounds IR dark regions, where several embedded sources (with H$-$K $\ge$ 2.3 mag), including massive stars IRS 1 and IRS 2 are present.\\ 
$\bullet$ The ALMA HNC(3--2) line data reveal a mirrored B-shaped feature (extent $\sim$19000 AU $\times$ 39000 AU) toward the eastern part of the molecular 
ring.  A significant velocity difference is observed between the outer portion and the base of the mirrored B-shaped feature, indicating molecular gas expansion at $\sim$2.25 km s$^{-1}$ toward this region of the molecular ring.\\
$\bullet$ The ALMA 1.14 mm continuum map shows an elongated filament-like feature (extent $\sim$0.24 pc). At its center, there is a small-scale dust shell-like feature (extent $\sim$0.04 pc $\times$ 0.07 pc; mass $\sim$7 M$_{\odot}$) hosting IRS~1. This shell-like feature is almost located toward the base of the B-shaped feature.\\
$\bullet$ One half of the small-scale dust shell-like feature contains a compact dust continuum source associated with molecular emission but lacking radio continuum emission. In contrast, the other half, which hosts four compact sources, shows no molecular emission but exhibits radio continuum emission.\\
$\bullet$ The study of the ALMA HNC(3--2) line data reveals distinct sub-structures in both the redshifted and blueshifted velocity components toward the B-shaped feature. These intersecting, braid-like substructures suggest instability in PDRs.\\
$\bullet$ The IR and dense molecular rings seem to be shaped by feedback from massive stars, driven by high pressure values (10$^{-8}$--10$^{-10}$ dynes cm$^{-2}$). \\

The findings in this paper consistently indicate that the Mon R2 HFS evolved from an 
IR-quiet to an IR-bright state, driven by the interplay between gas accretion and feedback from massive stars. 
The massive stars in the hub (i.e., IRS 1 and IRS 2) are responsible for the expansion of an H\,{\sc ii} region, causing the molecular ring to be pushed outward, while material continues to accrete into the system.
\section*{Acknowledgments}
We thank the reviewer for useful comments and suggestions, which greatly improved this manuscript. 
The research work at Physical Research Laboratory is funded by the Department of Space, Government of India. 
This paper makes use of the following ALMA data: ADS/JAO.ALMA\#2015.1.00453.S and ADS/JAO.ALMA\#2016.1.01144.S. ALMA is a partnership of ESO (representing its member states), NSF (USA) and NINS (Japan), together with NRC (Canada) , MOST and ASIAA (Taiwan), and KASI (Republic of Korea), in cooperation with the Republic of Chile. The Joint ALMA Observatory is operated by ESO, AUI/NRAO and NAOJ. The HST data presented in this article were obtained from the Mikulski Archive for Space Telescopes (MAST) at the Space Telescope Science Institute. The specific observations analyzed can be accessed via \dataset[10.17909/3ha3-wt08]{http://dx.doi.org/10.17909/3ha3-wt08}. 
This research made use of Astropy\footnote[6]{http://www.astropy.org}, a community developed core Python package for Astronomy \citep{astropy13,astropy18}. 
Figures were created using IDL software and matplotlib \citep{hunter07}.
%
%


\bibliographystyle{aasjournal}
\bibliography{reference}{}


\appendix
\restartappendixnumbering
%

\section{Feedback of massive stars in Mon~R2}
\label{s4sec:d}
%
%
IRS~1 has been suggested as the ionizing source responsible for the UCH\,{\sc ii} region (see Figure~\ref{fg1}a) and is classified as a B0 ZAMS star \citep{downes75,beckwith76,massi85,henning92,fuente10}.  
Hence, the impact of this massive star on its surroundings can be studied by analyzing various feedback pressure components exerted by the massive star, including the pressure of an H\,{\sc ii} region ($P_{\rm HII} = \mu m_{\rm H} c_{\rm s}^2\, \left(\sqrt{3N_\mathrm{UV}\over 4\pi\,\alpha_{\rm B}\, D_{\rm s}^3}\right)$), the radiation pressure ($P_{\rm rad}$ = $L_{\rm bol}/ 4\pi c D_{\rm s}^2$), and the stellar wind ram pressure ($P_{\rm wind}$ = $\dot{M}_{\rm w} V_{\rm w} / 4 \pi D_{\rm s}^2$) \citep[e.g.,][]{bressert12,dewangan17a}. 
One can notice that $P_{\rm HII}$ is proportional to $D_{\rm s}^{-3/2}$, while both $P_{\rm rad}$ and $P_{\rm wind}$ scale with $D_{\rm s}^{-2}$. 
Here, each pressure component as well as the total pressure ($P_{\rm total}$ = $P_{\rm HII}$ + $P_{\rm rad}$ + $P_{\rm wind}$) can be determined at the projected distance (i.e., $D_{\rm s}$) from the location of the massive star. In these equations, $N_\mathrm{UV}$ is the Lyman continuum photons \citep[i.e., 2.29 $\times$ 10$^{47}$ photons s$^{-1}$;][]{panagia73}, c$_{\rm s}$ is the sound speed of the photo-ionized gas \citep[i.e., 11 km s$^{-1}$;][]{bisbas09}, $\alpha_{\rm B}$ is the radiative recombination coefficient \citep[= 2.6 $\times$ 10$^{-13}$ $\times$ (10$^{4}$ K/$T_{\rm e}$)$^{0.7}$ cm$^{3}$ s$^{-1}$; see][]{kwan97}, $\mu$ is the mean molecular weight in the ionized gas
\citep[i.e., 0.678;][]{bisbas09}, m$_{\rm H}$ is the hydrogen atom mass, $\dot{M}_{\rm w}$ is the mass-loss rate 
\citep[i.e., 2.7 $\times$ 10$^{-9}$ M$_{\odot}$ yr$^{-1}$;][]{kobulnicky19}, 
$V_{\rm w}$ is the wind velocity of the ionizing source \citep[i.e., 1200 km s$^{-1}$;][]{kobulnicky19}, and L$_{\rm bol}$ is the bolometric luminosity of the 
source \citep[i.e., 2.5 $\times$ 10$^{4}$ L$_{\odot}$;][]{panagia73}. 

%
Considering $T_{\rm e}$ = 10$^{4}$~K, we obtained the values 
$P_{\rm HII}$ = [6.3 $\times$ 10$^{-9}$, 3.7 $\times$ 10$^{-9}$, 1.6 $\times$ 10$^{-10}$, 1.2 $\times$ 10$^{-10}$] dynes cm$^{-2}$; 
$P_{\rm rad}$  = [5.5 $\times$ 10$^{-9}$, 2.7 $\times$ 10$^{-9}$, 4.2 $\times$ 10$^{-11}$, 2.7 $\times$ 10$^{-11}$] dynes cm$^{-2}$; 
$P_{\rm wind}$ = [3.5 $\times$ 10$^{-11}$, 1.7 $\times$ 10$^{-11}$, 2.7 $\times$ 10$^{-13}$, 1.7 $\times$ 10$^{-13}$] dynes cm$^{-2}$; 
$P_{\rm total}$ = [1.2 $\times$ 10$^{-8}$, 6.4 $\times$ 10$^{-9}$, 2.0 $\times$ 10$^{-10}$, 1.4 $\times$ 10$^{-10}$] dynes cm$^{-2}$ at $D_{\rm s}$ = [0.07, 0.1, 0.8, 1] pc.  
The selected range of $D_{\rm s}$ spans small to moderately large distances around the Mon R2 hub, enabling a comprehensive assessment of the pressure distribution from the very close vicinity of IRS 1 and IRS 2 to the outer regions influenced by the filament convergence. Note that the filament convergence radius is reported to be $\sim$0.8 pc and the separation of IRS 1 and IRS 2 is $\sim$0.07 pc.  
%
The comparison of the different pressure values at various distances reveals that the ionized gas pressure 
($P_{\rm HII}$) consistently dominates over the other two components ($P_{\rm rad}$ and $P_{\rm wind}$) and  is the primary factor in determining the overall pressure distribution.    
%
$P_{\rm total}$ is dominated by $P_{\rm HII}$ at different $D_{\rm s}$ . 
It is noted that $P_{\rm total}$  or $P_{\rm HII}$ is also higher than the pressure of a typical cool molecular cloud ($P_{\rm MC}$) that ranges from approximately 10$^{-11}$ to 10$^{-12}$ dynes cm$^{-2}$, given a temperature of around 20 K and a particle density between $\sim$10$^{3}$ and 10$^{4}$ cm$^{-3}$ \citep[refer to Table 7.3 in][]{dyson80}. 
The pressure calculations are performed for the massive B0 ZAMS star, but the IR ring encloses two massive stars IRS~1 and IRS~2. 

Considering different pressure values, it is very much possible that the two massive stars have 
significantly influenced their immediate vicinity within $D_{\rm s}$ $<$ 1 pc. 
Therefore, the absence of dense molecular gas in the south-west portion of the molecular ring can be attributed to the impact of IRS 2 (see Figure~\ref{fg1}d). Similarly, the lack of dense gas in the direction of the dust shell-like feature is likely due to the influence of IRS 1.  In this context, comparing the immediate surroundings of IRS 1 in Figures~\ref{fg10}a and~\ref{fg10}c may provide valuable insights. 

\clearpage
%
%

%
%
\begin{table*}
\scriptsize
\setlength{\tabcolsep}{0.05in}
\centering
\caption{SHARC-II dust continuum sources at 350 $\mu$m (see squares in Figure~\ref{fg1}c). 
Table contains SHARC-II source designations, positions, major axis, minor axis, position angle, deconvolved angular size, integrated 
flux density (S$_{\nu}$), and clump mass (M$_{clump}$). All parameters, except for clump mass, are sourced from \citet{merello15}. 
Clumps located in the central/inner area (see the light coral contour in Figure~\ref{fg1}c) are indicated by daggers. 
Clump masses are computed using integrated fluxes, assuming a dust temperature of 23~K at a distance of 830 pc. 
For the remaining clumps, masses are estimated based on a dust temperature of 18.5~K (see text for additional details).} 
\label{tab1}
\begin{tabular}{lcccccccccccccc}
\hline  
  ID          & SHARC                  &  {\it l}     &   {\it b}      & major     & minor & Position angle  & Deconvolved  & S$_{\nu}$  & M$_{clump}$                   \\   
              &  Name                 & [degree]        &  [degree]    & axis ($''$)   &  axis ($''$)& [degree]        & angular size          ($''$)         &  [Jy]      &  (M$_{\odot}$)   \\   
\hline 						   	      	   							      
1 	      & G213.6578-12.6214     &   213.6572   &   $-$12.6215   &     3	 &    3  &   71   &   --   & 	 2.17  & 6.3 	 \\		    
2 	      & G213.6595-12.5817     &   213.6595   &   $-$12.5822   &    11	 &    5  &  159   &    15  & 	23.22  & 67.3	 \\
3 	      & G213.6606-12.5888     &   213.6613   &   $-$12.5902   &     5	 &    3  &  107   &   --   & 	 5.05  & 14.6	 \\
4 	      & G213.6608-12.6191     &   213.6603   &   $-$12.6193   &     5	 &    4  &   63   & 	 6 & 	 3.97  & 11.5	 \\
5 	      & G213.6676-12.6076     &   213.6680   &   $-$12.6071   &     6	 &    5  &   21   & 	10 & 	 8.63  & 25.0	  \\
6 	      & G213.6692-12.5962     &   213.6680   &   $-$12.5954   &     9	 &    3  &  105   &   --   & 	 3.47  & 10.1	 \\
7 	      & G213.6761-12.5986     &   213.6735   &   $-$12.5975   &     5	 &    4  &   74   & 	 5 &  	 3.22  & 9.3 	 \\
8 	      & G213.6769-12.5990     &   213.6773   &   $-$12.5986   &     3	 &    3  &  106   &   --   & 	 2.14  & 6.2 	 \\
9 	      & G213.6769-12.6064     &   213.6780   &   $-$12.6057   &    11	 &    6  &    4   & 	17 & 	26.34  & 76.3	 \\
10	      & G213.6827-12.6306     &   213.6817   &   $-$12.6312   &     6	 &    3  &    9   &   --   & 	 2.83  & 8.2 	 \\
11	      & G213.6864-12.6101     &   213.6861   &   $-$12.6107   &     6	 &    4  &    3   & 	 6 & 	 8.76  & 25.4	\\
12	      & G213.6872-12.6301     &   213.6864   &   $-$12.6306   &     5	 &    4  &   52   & 	 7 & 	 3.61  & 10.5	 \\
13$^{\dagger}$& G213.6942-12.6172     &   213.6929   &   $-$12.6219   &    14	 &    8  &  145   & 	23 &   140.50  &  246.1   \\
14$^{\dagger}$& G213.6946-12.5922     &   213.6909   &   $-$12.5923   &    18	 &    9  &   66   & 	28 &   251.36  &  440.3    \\
15	      & G213.6949-12.5729     &   213.6946   &   $-$12.5722   &     4	 &    2  &  144   &   --   & 	 1.08  &  3.1    \\
16$^{\dagger}$& G213.6967-12.6021     &   213.6945   &   $-$12.6011   &    13	 &    9  &  108   & 	25 &   237.45  &  416.0   \\		     
17$^{\dagger}$& G213.6979-12.6129     &   213.6980   &   $-$12.6137   &     8	 &    8  &  124   & 	17 &   142.31  &  249.3  \\
18	      & G213.6984-12.5801     &   213.6977   &   $-$12.5793   &     8	 &    4  &  125   & 	10 & 	 6.21  &  18.0   \\
19$^{\dagger}$& G213.7034-12.5909     &   213.7039   &   $-$12.5870   &    14	 &    9  &   33   & 	25 &   222.92  &  390.5  \\
20$^{\dagger}$& G213.7045-12.5989     &   213.7046   &   $-$12.5977   &    12	 &   11  &  100   & 	27 &   409.80  &  717.9   \\
21$^{\dagger}$& G213.7053-12.6095     &   213.7066   &   $-$12.6120   &     7	 &    5  &   67   & 	10 &   115.90  &  203.0  \\
22$^{\dagger}$& G213.7053-12.6170     &   213.7049   &   $-$12.6197   &    12	 &   10  &  157   & 	24 &   234.51  &  410.8  \\
23$^{\dagger}$& G213.7055-12.6054     &   213.7055   &   $-$12.6067   &     8	 &    7  &    5   & 	15 &   200.10  &  350.5 \\
24$^{\dagger}$& G213.7115-12.6162     &   213.7139   &   $-$12.6169   &    13	 &    7  &   66   & 	21 &   165.80  &  290.5 \\
25	      & G213.7142-12.6390     &   213.7136   &   $-$12.6393   &     4	 &    3  &  118   &   --   & 	 2.41  & 7.0  \\
26$^{\dagger}$& G213.7172-12.6125     &   213.7163   &   $-$12.6093   &    14	 &    8  &  157   & 	24 &   120.76  &  211.6  \\
27	      & G213.7180-12.6390     &   213.7178   &   $-$12.6391   &     8	 &    4  &  115   & 	 9 & 	 5.79  & 16.8   \\
28	      & G213.7204-12.5847     &   213.7202   &   $-$12.5852   &     8	 &    5  &  139   & 	12 & 	 7.30  &  21.2    \\
29	      & G213.7210-12.6004     &   213.7216   &   $-$12.5996   &     3	 &    2  &   98   &   --   & 	 2.44  &  7.1	 \\
30	      & G213.7271-12.6182     &   213.7275   &   $-$12.6188   &     4	 &    2  &  120   &   --   & 	 1.02  &  3.0	\\
31	      & G213.7355-12.6009     &   213.7358   &   $-$12.6008   &     4	 &    2  &  138   &   --   & 	 0.86  &  2.5	\\
32	      & G213.7378-12.6123     &   213.7358   &   $-$12.6125   &     9	 &    4  &   59   & 	 8 & 	 6.33  &  18.3   \\
33	      & G213.7389-12.6072     &   213.7391   &   $-$12.6063   &     9	 &    4  &  135   & 	 9 & 	 5.08  &  14.7    \\
34	      & G213.7424-12.6042     &   213.7421   &   $-$12.6036   &     5	 &    3  &  117   &   --   & 	 1.62  &  4.7    \\
\hline          
\end{tabular}
\end{table*}
\begin{table*}
\scriptsize
\setlength{\tabcolsep}{0.05in}
\centering
\caption{Physical parameters of ALMA continuum sources at 1.14 mm identified with {\it clumpfind} (see Section~\ref{subsec:data3a} for more details). 
Table contains IDs, positions, deconvolved FWHM$_{x}$ $\times$ FWHM$_{y}$, flux densities, and masses of the continuum sources. Masses are estimated using the integrated fluxes for a dust temperature = 23 K at a distance of 830 pc. The continuum sources ``t1--t5'', ``r1--r5'', ``c1--c5'', and ``u1--u5'' are labeled in Figure~\ref{fg4}c (see also Figures~\ref{fg4z}a and~\ref{fg4z}c).}
\label{tab2}
\begin{tabular}{lcccccccccccccc}
\hline  
 ID &  {\it l}  &   {\it b}	& FWHM$_{x}$ $\times$ FWHM$_{y}$ &  S$_{\nu}$  & M$_{s}$ 		  \\   
    & [degree]	&  [degree]	&              (arcsec$^{2}$)    &  [mJy]      & (M$_{\odot}$)   \\   
\hline 						   	      	   							      
 t1 & 213.70487 &  $-$12.59688 &   1.1 $\times$ 1.2 &	 126.31 &  1.0  \\
 t2 & 213.70458 &  $-$12.59742 &   4.3 $\times$ 2.5 &	  79.17 &  0.6  \\
 t3 & 213.70517 &  $-$12.59854 &   3.2 $\times$ 1.9 &	 101.31 &  0.8  \\
 t4 & 213.70550 &  $-$12.59962 &   1.8 $\times$ 2.6 &	  73.18 &  0.6  \\
 t5 & 213.70495 &  $-$12.60035 &   3.1 $\times$ 3.5 &	  92.98 &  0.7  \\
 r1 & 213.70802 &  $-$12.60399 &   3.0 $\times$ 3.3 &     38.02 &  0.3  \\ 		     
 r2 & 213.70547 &  $-$12.60540 &   4.1 $\times$ 8.6 &    882.12 &  7.0  \\ 		     
 r3 & 213.70106 &  $-$12.60313 &   2.8 $\times$ 3.9 &     36.46 &  0.3  \\ 		     
 r4 & 213.70029 &  $-$12.60414 &   1.8 $\times$ 1.8 &    194.29 &  1.5  \\ 		     
 r5 & 213.70151 &  $-$12.60605 &   1.1 $\times$ 2.4 &     21.10 &  0.2   \\		     
 c1 & 213.70547 &  $-$12.60540 &   2.8 $\times$ 3.9 &    369.31 &  2.9  \\
 c2 & 213.70558 &  $-$12.60356 &   2.1 $\times$ 1.8 &    106.52 &  0.8  \\
 c3 & 213.70502 &  $-$12.60428 &   2.5 $\times$ 2.4 &    206.01 &  1.6  \\
 c4 & 213.70624 &  $-$12.60472 &   1.5 $\times$ 3.2 &    121.11 &  1.0  \\
 c5 & 213.70521 &  $-$12.60295 &   2.4 $\times$ 2.6 &     69.54 &  0.5  \\
 u1 & 213.70635 &  $-$12.60876 &   1.8 $\times$ 2.7 &	 23.17 &  0.2  \\
 u2 & 213.70528 &  $-$12.60966 &   1.3 $\times$ 1.0 &	 14.32 &  0.1  \\
 u3 & 213.70635 &  $-$12.61053 &   1.2 $\times$ 4.4 &	 51.83 &  0.4  \\
 u4 & 213.70672 &  $-$12.61028 &   1.2 $\times$ 1.1 &	 20.31 &  0.2  \\
 u5 & 213.70732 &  $-$12.61172 &   2.1 $\times$ 1.9 &	 20.84 &  0.2  \\
\hline          
\end{tabular}
\end{table*}

\end{document}